\documentclass[a4paper]{article}%[a4paper,twocolumn,10pt]

\usepackage{amssymb}
\usepackage{graphicx}
\usepackage{amsmath}
\usepackage{mathtools}
\usepackage{times}
\usepackage{bm}
\usepackage{url}
\usepackage{multirow}
\usepackage{tabularx, ragged2e, booktabs}
\newcolumntype{Y}{>{\centering\arraybackslash}X}

\newcommand{\bfac }{} % *ac = *a that i want to clear
\newcommand{\bfb  }{} % *b new ones in *_07.tex
\newcommand{\bfc  }{} % *b new ones in *_07.tex
\newcommand{\bfd  }{} % *b new ones in *_07.tex
\newcommand{\bfe  }{} % *b new ones in *_07.tex

\begin{document}

\title{Rough basin boundaries in high dimension: Can we classify them experimentally?} % 17.10.19. new title because the negative LE is not really to do with predictability, and, the dim' of the unperturbed filamentary BB has something to do with predictability.  % Does the predictability of the second kind boil down to the predictability of the first kind?}
% V proposed to use pred of first and second kind in the title; he liked the intro discussing this.

\author{\large Tam\'as B\'odai and Valerio Lucarini \\
Centre for the Mathematics of Planet Earth, \\ Department of Mathematics and Statistics, University of Reading, UK} 

\maketitle

\begin{abstract}

We show that a known condition for having rough basin boundaries in bistable 2D maps holds for high-dimensional bistable systems that possess a unique nonattracting chaotic set embedded in their basin boundaries. The condition for roughness is that the cross-boundary Lyapunov exponent $\lambda_x$ {\bfac on the nonattracting set} is not the maximal one. Furthermore, we provide a formula for the generally noninteger co-dimension of the rough basin boundary, which can be viewed as a generalization of the Kantz-Grassberger formula. This co-dimension that can be at most unity can be thought of as a partial co-dimension, and, so, it can be matched with a Lyapunov exponent. We show {\bfac in 2D noninvertible- and 3D invertible minimal models,} that, formally, it cannot be matched with $\lambda_x$. Rather, the partial dimension $D_0^{(x)}$ that $\lambda_x$ is associated with in the case of rough boundaries is trivially unity. Further results hint that the latter holds also in higher dimensions. This is a peculiar feature of rough fractals. Yet, $D_0^{(x)}$ cannot be measured via the uncertainty exponent along a line that traverses the boundary. Indeed, one cannot determine whether the boundary is a rough or a filamentary fractal by measuring fractal dimensions. Instead, one needs to measure both the maximal and cross-boundary Lyapunov exponents numerically or experimentally.

\end{abstract}

\section{Introduction}\label{sec:Intro}

Beside chaotic attractors, nonattracting chaotic sets also have practical relevance~\cite{LT:2011}. They are associated with e.g. basins of attraction in multistabe systems, or, long-lived chaotic transients. Respective global properties quantified by characteristic numbers determine {\em predictability} and the {\em life time} of trajectories within a vicinity of the nonattracting set. Regarding nonattracting sets the most relevant concept of predictability is that of the second kind, concerning the {\em outcome} of an experiment in terms of the final state of the system out of a few alternatives~\cite{PhysRevE.87.042902}. This kind of predictability can be quantified by the {\em uncertainty exponent}~\cite{GREBOGI1983415,Tel_n_Gruiz:2006}, expressing the improvement of {\bfac predictability of} the outcome by determining the initial condition more precisely. To put this in to context, we note that attractors, on the other hand, are associated with predictability of the first kind only, when we are concerned with 
%the diminishing of the error of predicting some future state of the system as we improve the certainty 
% 13.01.2020.
{\bfc how the error in the prediction of the future state of the system changes as we change the precision in the definition}
of the initial conditions. 
{\bfe Error growth or decline in specific directions~\cite{Ginelli_2013} is measured by a spectrum of Lyapunov exponents (LE); and, concerning an initial condition with small {\em random} error, it is the {\em maximal} positive LE (MLE) that determines the asymptotic but infinitesimal error growth for trajectories confined to the invariant set that supports a measure.}
%This is quantified by the maximal Lyapunov exponent (MLE). %The
Lyapunov exponents %, which can be defined also for nonattracting invariant sets, 
are {\bfe thus} regarded to quantify {\em local} instabilities, while the {\em escape rate} $\kappa$, the inverse of the expected or {\em characteristic life time} of trajectories mentioned above, is regarded to quantify {\em global} instability. {\bfc LEs can be defined for both attractors and nonattracting invariant sets that support a measure, while attractors are clearly globally stable with no escape from them.~\cite{LT:2011}}

{\bfac Global characteristic numbers are in fact not completely independent. First,} the uncertainty exponent $\alpha$ has a straightforward one-to-one connection with the fractal dimension $D_{b,0}$ of the basin boundary, being simply the co-dimension $\alpha=D-D_{b,0}$, {\bfb where $D$ is the dimension of the phase space,} implying that the more space-filling the set, the poorer the predictability of the second kind. {\bfac Second,} a {\bfac connection (i)} of the predictability of the second kind and global instability, and, {\bfac third,} another {\bfac connection (ii)} of the global and local instabilities, can be given as follows. 
% 29.03.19. I stick here the eq.s first, as it's not a good practice to refere forward
We consider here discrete-time {\bfc bistable} systems that are possibly composed of coupling 
%a bistable ($X$) and a monostable ($Y$) 
{\bfc two subsystems}
(both of which can be multi-dimensional):
%, possessing two and one attractors, respectively:  

\begin{eqnarray}\label{eq:coupled_sys}
 X_{i+1}&=&f_X(X_i,Y_i;\epsilon_X,p),\label{eq:coupled_sys_X} \\
 Y_{i+1}&=&f_Y(X_i,Y_i;\epsilon_Y),\label{eq:coupled_sys_Y}
\end{eqnarray}
where the subsystems decouple for $\epsilon_{X}=\epsilon_{Y}=0$, such that %$f_X(X,Y;\epsilon_X=0)=f_X(X,Y=0;\epsilon_X)=\tilde{f}_X(X)$ and $f_Y(X,Y;\epsilon_Y=0)=f_Y(X=0,Y;\epsilon_Y)=\tilde{f}_Y(Y)$.
% 09.01.2020. cannot remember why i had it that way, but it's not true even for my minimal maps
$f_X(X,Y;\epsilon_X=0)=\tilde{f}_X(X)$ and $f_Y(X,Y;\epsilon_Y=0)=\tilde{f}_Y(Y)$.
{\bfc We assume that the unperturbed ($\epsilon_{X}=0$) subsystem (\ref{eq:coupled_sys_X}) has two co-existing attractors, %i.e., that it is a bistable system, 
while the unperturbed ($\epsilon_{Y}=0$) subsystem (\ref{eq:coupled_sys_Y}) has a unique globally attracting set. That is, the bistability of the coupled system derives from that of (\ref{eq:coupled_sys_X}).}
{\bfac Furthermore, we assume that a {\em single} nonattracting chaotic set is embedded in the basin boundary {\bfc of either (\ref{eq:coupled_sys_X}) ($\epsilon_{X}=0$) or the coupled system (\ref{eq:coupled_sys_X})-(\ref{eq:coupled_sys_Y})}, and so {\bfc they} possesses a {\em single} unstable dimension in which escape can occur. Such a nonattracting set (either a chaotic saddle or repeller) is said to be {\em low-dimensional}~\cite{LT:2011}. That is, the dimensionality of the nonattracting set in this sense is not to be confused with the dimensionality of the phase space of the system.} {\bfe Note that the stable manifold of the unique nonattracting set coincides with the basin boundary.} For {\bfac low-dimensional} nonattracting chaotic sets embedded in so-called {\em filamentary fractal boundaries}, ``locally consisting of a Cantor set
of smooth curves or surfaces''~\cite{LT:2011}, the Kantz-Grassberger relation~\cite{KG:1985} {\bfc holds}: 

\begin{equation}\label{eq:KG}
 \kappa=\lambda_x(1-D_1^{(x)}),\tag{KG}
\end{equation}
where $\lambda_x$ and $D_1^{(x)}$ are the cross-boundary LE {\bfb (i.e. the Lyapunov exponent describing the instability of the motion across the smooth boundary filament)} and associated~\cite{LT:2011} partial information dimension, respectively. {\bfc It} implies that 
\begin{description}
 \item[(i)] upon a change that leaves the LEs %(existing pre-perturbation $\epsilon_X=0$ under (i.b)) %or $\lambda_x$ 
(and so the predictability of the first kind) {\bfac practically} unchanged, a longer {\em characteristic life time}, i.e. a greater global stability, is implied by a poorer predictability of the second kind~\cite{PhysRevE.87.042902}, and vice-versa; and
 \item[(ii)] the global instability quantified by $\kappa$ is {\bfc in general} weaker than the local instability quantified by $\lambda_x$, simply because $0\leqq D_1^{(x)}\leqq1$, {\bfc as though the intricate folded geometry could trap the trajectory}.
\end{description}
As for (i) above the change may be (i.a) wrt. a parameter $p$, while $\epsilon_X=0$ in eq. (\ref{eq:coupled_sys_X}), or, (i.b) by introducing a perturbation, $\epsilon_X\neq0$. In the latter case it is meant that the basin boundary of the coupled system is still filamentary, and that the LEs of the uncoupled and coupled system are approximately the same. In the case of such a weak coupling, we use the term ``perturbation'' for $Y$ from the point of view of $X$.
{\bfe As shown by Wouters and Lucarini~\cite{Wouters_2012,Wouters2013}, switching on the coupling between the subsystems by setting $\epsilon_X\neq 0$ and/or $\epsilon_Y\neq 0$ can be formally treated as a perturbation using Ruelle's~\cite{Ruelle:2009} response theory.}

As discussed in Sec. \ref{sec:dim_formulae} {\bfac in the following}, for another type of boundary, a {\em continuous fractal boundary}, which is rough, nowhere differentiable, KG does not apply, but rather a more generic formula (\ref{eq:uncert}), in which $\lambda_x$ and $(1-D_1^{(x)})$ are replaced by the MLE $\lambda_{max}$ and the co-dimension of the boundary, i.e. $\alpha$, respectively. 
% 09.01.2020.
{\bfb This implies completely new properties, including the generic invalidity of (i) and (ii), which will be further discussed in Sec.~\ref{sec:discuss}.}
{\bfb It becomes thus} possible to have a very poor predictability of the second kind due to a strongly space-filling boundary, while the trajectory life time is fairly short, {\bfb with a global instability being about the same as the local cross-boundary instability}. The obvious -- and perhaps very common -- way that this situation can arise is that we have a bistable system ($X$) that is perturbed by another system ($Y$) whose behaviour is noise-like~\footnote{That is, the characteristic time scales of $Y$ are much smaller than that of $X$.} compared to that of the bistable one. A weak enough noise-like perturbation does not alter the global instability \cite{Franaszek:1991} but renders the outcome completely unpredictable (although the thickness of the boundary scales with the perturbation strength). {\bfac That is, the poor predictability of the first kind of the fast system $Y$ will impact the predictability of the second kind of the coupled system. In the extreme, predictability of the second kind is completely lost, $\alpha=0$, with an extreme time-scale separation $\lambda_x/\lambda_{max}\rightarrow0$, even if in the uncoupled/unperturbed ($\epsilon_X=0$) bistable system $X$ we had perfect predictability of the second kind, $D_1^{(x)}=0$, $\alpha=1$.}

{\bfb We encountered} such a situations %has been encountered  
in the case of a bistable climate model of intermediate complexity~\cite{0951-7715-30-7-R32}. In this model the so-called snowball-snow free bistability is created by the ice-albedo positive feedback. This effect can be modeled by very simple 0-D energy balace models (EBM) {\bfb (where ``0'' in ``0-D'' refers to the spatial extension of variables)}. In our model {\bfb studied in~\cite{0951-7715-30-7-R32}} a 1-D diffusive heat equation serves the same purpose, which, however, has in common with the 0-D EBM a nonchaotic solution. When this ocean-ice model component ($X$) was coupled with the chaotic atmosphere ($Y$), the originally regular basin boundary was found to turn into a practically space-filling object. 

{\bfc Attaining our motivating objective,} we {\bfb are able to} claim here that what we encountered was in fact a rough continuous very thick fractal, because, first, the condition $\lambda_x<\lambda_{max}$ found for 2D maps~\cite{GOY:1983,VSE:2009} we argue {\bfb in Sec. \ref{sec:condition}} to be applicable to high-dimensional systems, and, second, the condition was in fact satisfied by the coupled climate model. As for {\bfb the second point}, we did not measure the cross-boundary LE of the coupled/perturbed climate model, only the MLE, but we did measure $\lambda_x$ of the 1-D EBM in~\cite{Bodai2015}, and clearly it is not altered significantly by a rather weak coupling and the weakened diffusivity {\bfb applied in~\cite{0951-7715-30-7-R32}}, while $\lambda_x$ and $\lambda_{max}$ are obviously vastly different {\bfb representing climatic and weather processes, respectively}. 

Next, in Sec. \ref{sec:theory} we reproduce the condition for roughness in a general setting, and also provide dimension formulae for rough boundaries. In Sec. \ref{sec:model} we provide minimal models that feature rough boundaries, among them a prototypical model for a new kind of mixed filamentary-rough boundary. In Sec. \ref{sec:dim_numerics} we report on our numerical computations performed to determine the fractal dimension of the boundary. Finally, in Sec. \ref{sec:discuss} we discuss our results here and those in~\cite{0951-7715-30-7-R32}, and pose some open questions of geophysical relevance.

\section{Theory}\label{sec:theory}

\subsection{Condition for roughness}\label{sec:condition}

Grebogi et al.~\cite{GOY:1983} provided for the first time a condition for the roughness of the basin boundary in a 2D map when this rough boundary can be described as a Weierstarss function in terms of a Fourier series whose derivative is a nonconvergent series. The latter can be viewed as a recipe for creating a map with a rough basin boundary. However, Vollmer et al.~\cite{VSE:2009} derives the same condition in an alternative way, not requiring the boundary to be described by a Weierstrass function. This is what we reproduce next, pointing out in addition that 1) it applies not only to 2D maps but to any high-dimensional one, and 2) also to systems with a fractal boundary which is filamentary when $\lambda_x=\lambda_{max}$. %the condition for roughness is not satisfied. 
% 10.01.2020. Since i deleted the text about $D_1^{(x)}=1$ from the Intro. i need to comment the follwoing out :(
%Regarding 2), a parameter $p$ of (\ref{eq:coupled_sys_X})-(\ref{eq:coupled_sys_Y}) may be changed in a way that a critical value $p_*$ at $\lambda_x=\lambda_{max}$ is crossed, so that $D_1^{(x)}>0$ when $p<p_*$, and $\lambda_x<\lambda_{max}$ and so {\bfa possibly} $D_1^{(x)}=1$ when $p>p_*$. 

Assume that 
%a coordinate transform has been performed with which $x$ denotes the variable transversal to the boundary and $y$ denotes a variable describing motions within the boundary. That is, the coordinate transformation rectifies the basin boundary. Note that the boundary is first assumed to be regular, when it can actually be rectified. 
$x$ and $y$ denote some local coordinates that describe motions in a coarse sense ``across'' and %within 
``along'' a basin boundary, respectively, of a discrete-time dynamical system. 
Therefore, the small perturbations around the boundary evolve as:

\begin{eqnarray}
 \delta x_n &=& \prod_{i=1}^n\Lambda_{x,i}\delta x_0 + \epsilon\sum_{j=0}^{n-1}\prod_{i=1}^j\Lambda_{x,i}\delta y_{n-1-j},  \label{eq:pert_dyn_1} \\
 \delta y_n &=& \prod_{i=1}^n\Lambda_{y,i}\delta y_0, \label{eq:pert_dyn_2}
\end{eqnarray}
where the $\Lambda$'s are local Lyapunov numbers. Note that eq. (\ref{eq:pert_dyn_1}) derives from the linear evolution equation (being a recursive formula) 

\begin{equation}\label{eq:recursive}
\delta x_{i+1} = \Lambda_{x,i}\delta x_i + \epsilon\delta y_i;
\end{equation} 
and from eq. (\ref{eq:pert_dyn_2}) the $x$-dynamics is readily transformed out, having utilized the fact that the dynamics is {\em constrained} to a surface being the basin boundary. Note also that a time-independent constant $\epsilon$ means that, to start with, we consider the case of {\em additive} perturbation, i.e., $f_x(x,y;\epsilon_X=\epsilon)=\tilde{f}_x(x)+\epsilon y$ in eq. (\ref{eq:coupled_sys_X}) simplified to a 2D situation. In eq. (\ref{eq:pert_dyn_1}) we express $\delta y_{n-1-j}$ using eq. (\ref{eq:pert_dyn_2}) and rearrange it as:

\begin{equation}\label{eq:grad_nonconst}
 \frac{\delta x_0}{\delta y_0} = -\epsilon \frac{\sum_{j=0}^{n-1}\prod_{i=1}^j\Lambda_{x,i} \prod_{i=1}^{n-1-j}\Lambda_{y,i}}{\prod_{i=1}^n\Lambda_{x,i}},
\end{equation} 
which owes to the fact that $\delta x_n$ is bounded when the perturbation is chosen in a special way that the perturbed trajectory stays on the boundary. That is, the degree-of-freedom of choosing such a perturbation is one, not two. With constant local Lyapunov numbers this equation would simplify to:

\begin{eqnarray}
 \frac{\delta x_0}{\delta y_0} = -\epsilon \frac{\sum_{j=0}^{n-1}\Lambda_{x}^j \Lambda_{y}^{n-1-j}}{\Lambda_{x}^n} = -\frac{\epsilon}{\Lambda_x}\sum_{j=0}^{n-1}\left(\frac{\Lambda_y}{\Lambda_x}\right)^{n-1-j}=-\frac{\epsilon}{\Lambda_x}\sum_{i=0}^{n-1}r^i,\label{eq:grad_const} \\ 
 r = \frac{\Lambda_y}{\Lambda_x}.
\end{eqnarray} 
A finite value for the left hand side would mean in the limit of $n\rightarrow\infty$ that the boundary is locally smooth.
%rectification -- our assumption -- was actually possible. 
It turns out that it is only possible if the Lyapunov number or exponent is smaller across the boundary than the other one ($r<1$), yielding a convergent series. If not ($r>1$), the boundary is not smooth locally  
%regular and so not rectifiable (reductio ad absurdum), 
but can be viewed as rough, not %where 
differentiable. When $\Lambda_{x,i}$, $\Lambda_{y,i}$ do vary over the nonattracting set, and so in time along a trajectory, in terms of $j$ appearing in eq. (\ref{eq:grad_nonconst}), these are the smaller values of $j$ that need to be kept in check, corresponding to larger powers of some $r$ of the time-independent formula (\ref{eq:grad_const}), and so in the limit, these are indeed the {\em average} Lyapunov exponents, as arithmetic averages of $\ln\Lambda_{x,y,i}$'s, that determine whether the series is convergent. {\bfb That is,}

\begin{align}
&\lim_{n\to\infty}\frac{1}{\sum_{j=0}^{n-1}\left(\frac{\Lambda_y}{\Lambda_x}\right)^{n-1-j}}=0\Rightarrow\\&\lim_{n\to\infty}\frac{1}{\sum_{j=0}^{n-1} \frac{\prod_{i=1}^{n-1-j}\Lambda_{y,i}}{\prod_{i=1}^{n-1-j}\Lambda_{x,i}}}=0\Rightarrow\\&\lim_{n\to\infty}\frac{\prod_{i=1}^n\Lambda_{x,i}}{\sum_{j=0}^{n-1}\prod_{i=1}^j\Lambda_{x,i} \prod_{i=1}^{n-1-j}\Lambda_{y,i}}=0.
\end{align}
(We can denote the (geometric) average Lyapunov numbers simply by $\Lambda_{x}$, $\Lambda_{y}$, just like the constant local Lyapunov numbers in the special case.) Besides that, we note that a spatial dependence of $\delta x_0/\delta y_0$ arises from the finite terms of the series. We also point out that 
%eq. (\ref{eq:recursive}) assumes an additive coupling resulting in a constant $\epsilon$. 
when the coupling is also nonlinear, such that 
%$x_{i+1}=f(x_i,y_i)$
the generic form of eq. (\ref{eq:coupled_sys_X}) applies, which results in a time-dependent $\epsilon_i$ (to replace the constant in eq. (\ref{eq:recursive})), the condition on roughness is {\em unchanged}. % for the same reason as with the time-dependent $\Lambda_{x,i}$, $\Lambda_{y,i}$: blah blah.
Next, we point out two more possible generalisations.

1) Consider a multi-dimensional boundary, with state variables $y_d$ %within 
``along'' this hyper-surface, $d=1,\dots,D-1$, $D$ being the phase space dimension. Then, eqs. (\ref{eq:pert_dyn_1}) and (\ref{eq:pert_dyn_2}) generalise straightforwardly as:

\begin{eqnarray}
 \delta x_n &=& \prod_{i=1}^n\Lambda_{x,i}\delta x_0 + \sum_{d=1}^{D-1}\epsilon_d\sum_{j=0}^{n-1}\prod_{i=1}^j\Lambda_{x,i}\delta y_{d,n-1-j},  \label{eq:pert_dyn_1_gen} \\
 \delta y_{d,n} &\sim& \Lambda_{y}^n, \label{eq:pert_dyn_2_gen}
\end{eqnarray}
where we retain the simplified notation $\Lambda_{y}$ for the maximal one out of all $\Lambda_{y,d}$. %belonging to the boundary. 
Note that in eq. (\ref{eq:pert_dyn_2_gen}) we only state the asymptotic behaviour, indicated by the symbol $\sim$, which also means that we suppress the indication of a constant of proportionality. We also point out that it is the maximal $\Lambda_{y}$ indeed that appears in all $D-1$ components of the subsystem (\ref{eq:pert_dyn_2_gen}); the components differ only wrt. the suppressed constant of proportionality. Therefore, we can express a directional derivative by rearranging (\ref{eq:pert_dyn_1_gen}) for $\delta x_0$ and dividing it by a normalised linear combination $\sum_d^{D-1}c_d\delta y_{d,0}$, where $\sum_d^{D-1}c_d=1$. This will be finite again, clearly, if $r<1$. It should be noted that eqs. (\ref{eq:pert_dyn_2_gen}) apply generically, i.e., with probability one, unless a direction is taken for the directional derivative that aligns with a covariant Lyapunov vector (CLV)  %\cite{} 
of the system~\cite{Ginelli_2013}. In that case eqs. (\ref{eq:pert_dyn_2_gen}) modify to be

\begin{equation}\label{eq:pert_dyn_2_clv}
 \delta y_{d,n} \sim \Lambda_{y,d^*}^n,
\end{equation} 
where $\Lambda_{y,d^*}$ can be any one of {\bfac the positive} $\Lambda_{y,d}$'s, %$d=1,\dots,D-1$, 
belonging to the CLV in question, the same value applying to all components of (\ref{eq:pert_dyn_2_clv}). In this case it is possible that the derivative exists given that $\Lambda_{y,d^*}/\Lambda_{x}<1$ while we have a rough surface because $r>1$, {\bfac i.e., the surface is not necessarily rough in every direction}.

2) Roughness can be regarded as a local property because the derivative belongs to a particular locale on the boundary. The neighbourhood of such a locale can be characterized %-- in order to establish the assumption -- 
the same way in the case of a filamentary fractal boundary and a regular boundary. On the other hand, roughness should be considered as a global property too because if the derivative does not exists in one locale, neither does it exist in any other one, given that the condition is based on global properties, the average Lyapunov numbers/exponents. Therefore, 
{\bfc also a filamentary fractal boundary should turn rough upon some change that changes $\lambda_x=\lambda_{max}$ to $\lambda_x<\lambda_{max}$. This change may be brought about either by changing $p$ of the coupled system (\ref{eq:coupled_sys_X})-(\ref{eq:coupled_sys_Y}) from some $p_1$ to $p_2$, or, by switching on the coupling going from $\epsilon_X=0$ to $\epsilon_X\neq0$. However,}
%the condition of roughness should apply also to boundaries that are filamentary fractals when the condition for roughness is not satisfied. When the condition is satisfied, 
the result is a new type of rough boundary. %OK, BUT HOW DO YOU DEFINE WHAT IS NEW ABOUT IT? 

\subsection{Dimension formulae}\label{sec:dim_formulae}

\subsubsection{Co-dimension of the basin boundary}

In order to establish dimension formulae applying %specifically % 10.01.2020. by this i meant that $D_1^{(x)} = 1$ but the generalisation of (KG) is ALSO applicable to filamentary bb's 
to rough boundaries, we invoke the following generic relations {\bfac established in~\cite{PhysRevE.54.4819}. The information dimensions of the unstable and stable manifolds of a nonattracting invariant set, respectively, are:
\begin{eqnarray}
 D_{u,1} = U + I + \frac{K_1 - \sum_{i=1}^I\lambda_i^-}{\lambda_{I+1}^-},\label{eq:Du1} \\
 D_{s,1} = S + J + \frac{K_1 - \sum_{j=1}^J\lambda_j^+}{\lambda_{J+1}^-},\label{eq:Ds1}
\end{eqnarray}
where $U$ ($S$) is the number of positive $\lambda_i^+$ (negative $\lambda_i^-$) Lyapunov exponents ($U+S=D$, $-\lambda_{S}^- \leq -\lambda_{S-1}^- \leq \dots \leq -\lambda_{1}^- \leq 0 \leq \lambda_{1}^+ \leq \dots \leq \lambda_{U-1}^+ \leq \lambda_{U}^+$), $I$ ($J$) is the largest integer for which the numerator of the fraction in eq. (\ref{eq:Du1}) ((\ref{eq:Ds1})) is still positive, and 
\begin{equation}\label{eq:K1}
 K_1 = \sum_{j=1}^U\lambda_j^+D_1^{(j)}
\end{equation}
is the so-called metric entropy. In the latter $D_1^{(j)}$ are the partial dimensions belonging to the LEs $\lambda_j^+$, while partial dimensions $D_1^{(i)}$ also belong to the negative LEs $\lambda_i^-$. These %certainly 
sum up to the full dimension of the invariant set: 
\begin{equation}\label{eq:dimsum1}
 D_1 = \sum_{i,j}D_1^{(i,j)}.
\end{equation}
Otherwise, this dimension is
\begin{equation}\label{eq:dimsum2}
 D_1 = D_{u,1} + D_{s,1} - D,
\end{equation}
because the nonattracting set is the intersection set of its stable and unstable manifolds, {\bfb and, in general~\cite{Mandelbrot1989}, 
% VL gave me this reference in Paris: https://books.google.co.kr/books?id=_h4GCAAAQBAJ&pg=PA3&lpg=PA3&dq=negative+dimensions+and+introduction+to+multifractals&source=bl&ots=YvL1KmuAV2&sig=ACfU3U0z5CIFt-KHUPWOHogTamQpIqUonQ&hl=en&sa=X&ved=2ahUKEwj7uKyA0PXmAhXMaN4KHU_LA-oQ6AEwBHoECAoQAQ#v=onepage&q=negative%20dimensions%20and%20introduction%20to%20multifractals&f=false
%
% along with: https://users.math.yale.edu/public_html/People/frame/Fractals/FracAndDim/DimAlg/intersection.html
the co-dimension of the intersection set $S$ is the sum of the co-dimensions of the two sets $S_1$ and $S_2$}: 

\begin{equation}\label{eq:dim_intersect}
	D-D_{S}=D-D_{S_1} + D-D_{S_2}.
\end{equation}
Escape takes place along directions in which the partial dimensions are less than maximal, as expressed by the following:
\begin{equation}\label{eq:kappa_sum}
 \kappa = \sum_{j=1}^U\lambda_j^+(1-D_1^{(j)}).
\end{equation}
For invertible systems only:
\begin{equation}\label{eq:info_conserv}
 \sum_{j=1}^U\lambda_j^+D_1^{(j)} = \sum_{i=1}^S\lambda_i^-D_1^{(i)}.
\end{equation}
Eqs. (\ref{eq:Du1},\ref{eq:Ds1},\ref{eq:K1},\ref{eq:kappa_sum},\ref{eq:info_conserv}) above are Eqs. (8.21,8.24,8.8,8.7,8.10) of~\cite{LT:2011}, respectively. 
% 16.01.2020. moved this bit here and with Next, we indicate i start a new subsection -- teazed out by V.
{\bfe Note that (KG) is only applicable to regular or filamentary fractal boundaries when $\lambda_x=\lambda_U$, and it derives rather straightforwardly from eq. (\ref{eq:kappa_sum}) as follows. Even if there was chaotic dynamics {\em within} the boundary, it would be a so-called {\em relative attractor} in that {\em smooth sub}space with no escape from it. This means that $D_j^{(j)}=1$, $j<U$, which leaves $D_1^{(x)}=D_1^{(U)}$ as the only nontrivial partial dimension belonging to a positive LE $\lambda_j^+=\lambda_x=\lambda_U$, leaving a single term of the sum in eq. (\ref{eq:kappa_sum}).} 
{\bfe Nevertheless, when the perturbations are {\em weak}, an extra equation (approximation, in fact) can be obtained observing that}
%A final equation is obtained from our observation that with {\em weak} perturbations 
$\kappa\approx\tilde{\kappa}$ and $\lambda_x\approx\tilde{\lambda}_x$, where the tilde specifies quantities belonging to the unperturbed ($\epsilon_X=0$) system, which are related by (KG); therefore:
\begin{equation}\label{eq:kappa_approx}
 \kappa\approx\lambda_x(1-\tilde{D}_1^{(x)}).
\end{equation}
Note that $\lambda_x$ is one of the $\lambda_j^+$'s. %, {\bfe and we can have this extra equation even if we don't know which $D_1^{(j)}<1$ in eq. (\ref{eq:kappa_sum})}. % Actually, i don't like this comment because we have the extra eq. because of the small perturbation and having that we will actually know which $D_1^{(j)}<1$.
%{\bfe We point out that, in contrast with eq.~\ref{eq:kappa_sum}, our extra ``equation'' (approximation, in fact) involves only the single $\lambda_x$ that is one of all $\lambda_j^+$'s.}

We can {\bfe now} obtain a formula for the dimension $D_{b,1}$, {\bfe applicable also to} a rough basin boundary {\bfe in the weak perturbation limit}, making use of the following two facts. First, eqs. (\ref{eq:K1}) and (\ref{eq:kappa_sum}) can be combined to yield $K_1 = \sum_{j=1}^U\lambda_j^+ - \kappa$, which can be substituted in the fraction of eq. (\ref{eq:Ds1}). Second, eq. (\ref{eq:kappa_approx}) and roughness $\lambda_x<\lambda_U$ implies that $\kappa<\lambda_U$, and, therefore, $J=U-1$ in eq. (\ref{eq:Ds1}). Bearing in mind that the basin boundary is identical to the stable manifold of the unique nonattracting set in it{\bfd ---just like the stable manifold of an attractor is space filling~\cite{LT:2011}---}we have:

\begin{equation} \label{eq:uncert}
 \alpha = D - D_{b,0} \approx D - D_{b,1} = D - D_{s,1} = \kappa/\lambda_{U}.
\end{equation} 
As we have already indicated, (\ref{eq:uncert}) is our generalisation of (KG), {\bfe such that (KG) is recovered when $\lambda_U=\lambda_x$ in the case of a filamentary fractal basin boundary. See eq. (5.13) of Ref.~\cite{LT:2011}, identical with our eq. (\ref{eq:uncert}), which was derived for a 2D noninvertible map.} %{\bfb
%applicable also to rough boundaries. 

% Note that (KG) is} only applicable to regular or filamentary fractal boundaries when $\lambda_x=\lambda_U$, {\bfb and it derives rather straightforwardly from eq. (\ref{eq:kappa_sum}) as follows.} %and
% Even if there was chaotic dynamics {\em within} the boundary, it would be a so-called {\em relative attractor} in that {\bfb smooth} {\em sub}space with no escape from it. {\bfb This means that} %, that is,
% $D_j^{(j)}=1$, $j<U$, %leaving 
% {\bfb which leaves} $D_1^{(x)}=D_1^{(U)}$ as the only nontrivial partial dimension belonging to a positive LE $\lambda_j^+=\lambda_x=\lambda_U$, {\bfb leaving a single term of the sum in eq. (\ref{eq:kappa_sum}). % as determined by (KG).

\subsubsection{Partial dimensions of rough basin boundaries}

Next, we indicate that for rough boundaries $D_1^{(x)}$ becomes trivially unity, which is a rather counterintuitive new characteristic {\bfe conflicting with (KG)}, as it suggests---considering eq. (\ref{eq:kappa_sum})---that escape does not take place in the cross-boundary direction. 
%This paradox should have to do with the fact 
{\bfc Although we remark} that the cross-boundary direction can actually not be defined for a rough boundary.} 

%Next, 
Consider a 2D noninvertible map with two positive LEs featuring a rough boundary  $\lambda_x<\lambda_y=\lambda_{U}$. Since there are no negative LEs, the fraction in eq. (\ref{eq:Du1}) disappears and $I=0$, yielding a full-dimensional unstable manifold $D_{u,1}=D=2$; {\bfac and also that the nonatracting set is identical with the whole basin boundary, $D_{s,1}=D_1$. (See also Sec. 8.3.1 of~\cite{LT:2011} treating the model in~\cite{GOY:1983}.)} Furthermore, eqs. (\ref{eq:dimsum1},\ref{eq:dimsum2},\ref{eq:kappa_sum},\ref{eq:uncert}) imply that
\begin{eqnarray}
 D_1^{(x)} &=& 1, \label{eq:part_dim_x} \\
 D_1^{(y)} &=& 1-\frac{\lambda_x}{\lambda_{y}}(1-\tilde{D}_1^{(x)}). \label{eq:part_dim_y}
\end{eqnarray}
We %emphasize 
{\bfb reiterate} that $D_{u,1}=2$, $D_{s,1}=D_1$ and $D_1^{(x)} = 1$ are completely new features of rough fractals with respect to filemanetray fractals. 
In the case of a regular unperturbed boundary $\tilde{D}_1^{(x)}=0$, and so 
\begin{equation} \label{eq:dim_pred_reg}
 D_1^{(y)} \approx 1-\frac{\lambda_x}{\lambda_{y}},
\end{equation} 
that is, the dimension can be predicted just by the Lyapunov exponents. In the case of a filamentary fractal unperturbed boundary we have to know the nontrivial fractional $\tilde{D}_1^{(x)}$ too. %This answers the question of the title. 
We can take the point of view that $D_1^{(y)}=\tilde{D}_1^{(x)}+\hat{D}_1$ is made up of a {\em contribution} from the filamentary fractality of the unperturbed boundary and a contribution from roughening, respectively, as if the contributions were partial dimensions. This implies that

\begin{equation} \label{eq:dim_rough}
 \hat{D}_1 = (1-\frac{\lambda_x}{\lambda_{y}})(1-\tilde{D}_1^{(x)}).
\end{equation} 
That is, even if the perturbation is noise-like, i.e. $\lambda_x\ll\lambda_{y}$, the contribution from roughening can be at most $(1-\tilde{D}_1^{(x)})$, which achieves the maximally possible $D_1^{(y)}=1$. I.e., the contribution of roughening is not independent of $\tilde{D}_1^{(x)}$, and so the latter cannot be considered a partial dimension on its own.

{\bfac Next, consider a 3D invertible map with two positive LEs and one negative LE, featuring a rough boundary $\lambda_x<\lambda_y=\lambda_{U}$. We have one more unknown sought-for variable $D_1^{(z)}$ with respect to the noninvertible 2D case, but we have one more equation as well: it is eq. (\ref{eq:info_conserv}). We can obtain (save the algebraic manipulation), therefore, that eqs. (\ref{eq:part_dim_x})-(\ref{eq:part_dim_y}) hold, and we have a further nontrivial partial dimension:
\begin{equation}
 D_1^{(z)} = \frac{\lambda_y + \lambda_x\tilde{D}_1^{(x)}}{\lambda_z}.
\end{equation}
Note that since $\lambda_y + \lambda_x<\lambda_z$ in the invertible dissipative system and $0\leq\tilde{D}_1^{(x)}\leq1$, we have $D_1^{(z)}<1$, indeed. We emphasize that the seemingly paradoxical $D_1^{(x)} = 1$ can hold even in invertible, possibly physical systems. 

Going further, in high dimensions, the partial dimensions seem to be under-determined by the available system of three linear equations:
\begin{eqnarray}
 \sum_{i,j}D_1^{(i,j)} &=& -\frac{\kappa}{\lambda_U} + U + I + \frac{\sum_{j=1}^U\lambda_j^+ - \kappa - \sum_{i=1}^I\lambda_i^-}{\lambda_{I+1}^-}, \label{eq:sys4dims1} \\
 \sum_{j=1}^U\lambda_j^+D_1^{(j)} &=& \sum_{j=1}^U\lambda_j^+ - \kappa, \label{eq:sys4dims2}\\
 \sum_{j=1}^U\lambda_j^+D_1^{(j)} &-& \sum_{i=1}^S\lambda_i^-D_1^{(i)} = 0, \label{eq:sys4dims3}
\end{eqnarray}
where $\kappa$ is also given in terms of a LE as per eq. (\ref{eq:kappa_approx}). In $D=4$, we can easily parametrize the three remaining partial dimensions by possible values of a selected one. Let us consider the example of the ``folded-towel'' map of R\"ossler~\cite{ROSSLER1979155}, whose LEs are: $\lambda_2^+= 0.430,\ \lambda_3^+= 0.377, \lambda_1^-= 3.299$ and choose $\lambda_1^+=\lambda_x=0.2$. Fig. \ref{fig:under_det_dim} shows the ``possible'' values of the partial dimensions. However, the only possible value for $D_1^{(x)}$ is 1, given the contraint $0\leq D_1^{(i,j)}\leq 1$. This prompts that we cannot exclude the possibility that $D_1^{(x)}=1$ in any higher dimension.

\begin{figure*} [t!]
    \begin{center}
            \scalebox{0.5}{\includegraphics{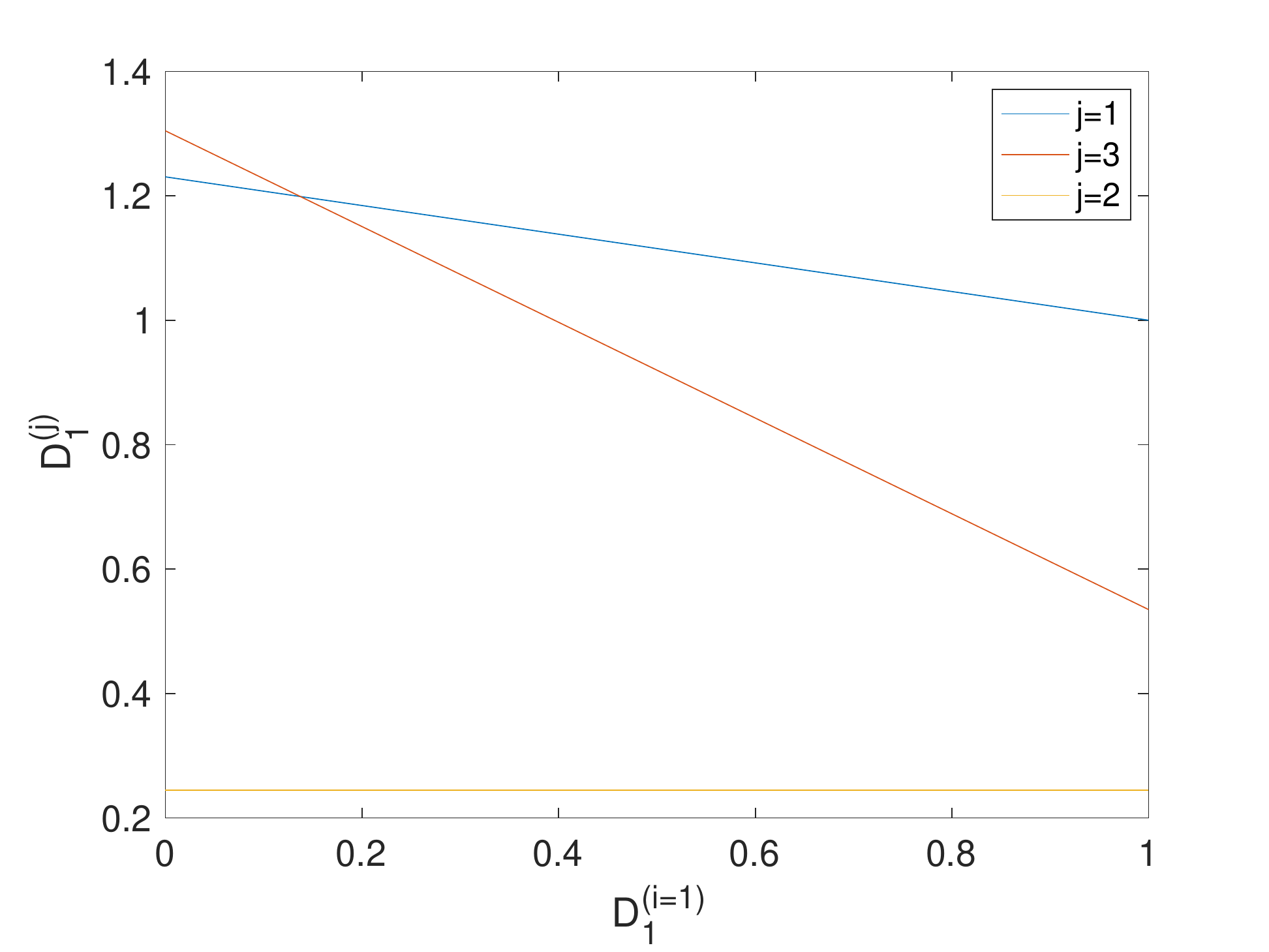}} 
        \caption{\label{fig:under_det_dim} 
         Parametric solution of eqs. (\ref{eq:sys4dims1},\ref{eq:sys4dims2},\ref{eq:sys4dims3}) for ``possible'' values of the partial dimensions of the ``folded-towel'' map coupled with a bistable linear system. See details in the main text.}
    \end{center}
\end{figure*}

}

\section{Minimal models for rough basin boundaries}\label{sec:model}

{\bfc In order to clarify some of the previous findings, we present here some mathematical toy models of interest.}

\subsection{Simple rough boundary}

A minimal model for bistable systems with `uncomplicated' cross-boundary and `complicated' in-boundary dynamics can be constructed as follows: 

\begin{eqnarray}
 x_{n+1} &=& ax_n + d(y_n - 1/2),  \label{eq:lin_1} \\
 y_{n+1} &=& \mu\min(y_n,1-y_n),\ \mu = 1 + \exp(c|x_n|). \label{eq:tent} 
\end{eqnarray} 
Loosely speaking, the cross-boundary dynamics is governed (dominated) by a linear equation (\ref{eq:lin_1}), and the in-boundary dynamics is a chaotic process produced by a noninvertible 1D map, the tent map (\ref{eq:tent}) in this example. Since the tent map is noninvertible, the whole 2D ($D=2$) system is noninvertible. The coupling is bidirectional. The in-to-cross-boundary coupling is additive, of strength $d$, and the cross-to-in-boundary coupling is multiplicative/parametric through $\mu(x)$. The latter coupling is not needed for a truly minimal model that features a rough basin boundary. The model with $c<0$ can be considered a minimal model for a rough boundary that cannot be described by a (Weierstrass) function $W(y)$. %This is also to say that the model coordinates $x$, $y$ do not correspond exactly to cross- and in-boundary directions. However, we continue to denote the cross- and in-boundary LEs by $\lambda_x$ and $\lambda_y$,  respectively. 
The bistability is between two attractors at $\pm\infty$ wrt. $x$. For our exercise we chose $a=1.1$, %(CHANGED FROM $1+a$ to $a$, and from $b$ to $d$!!!) 
$d=0.01$ and $c=-1$.

Given that the perturbation is weak, the LEs of the nonattracting set are approximately those of the unstable fixed point and attractor of the uncoupled bistable and chaotic subsystems, respectively:

\begin{equation}\label{eq:LEs}
 \lambda_x\approx\ln a, \quad \lambda_y\approx\ln2.
\end{equation} 
Note that small values of $x$ correspond to the nonattracting set embedded in the boundary, and so $\mu\approx2$ on the nonattracting set.

\begin{figure*} [t!]
    \begin{center}
            \scalebox{0.25}{\includegraphics{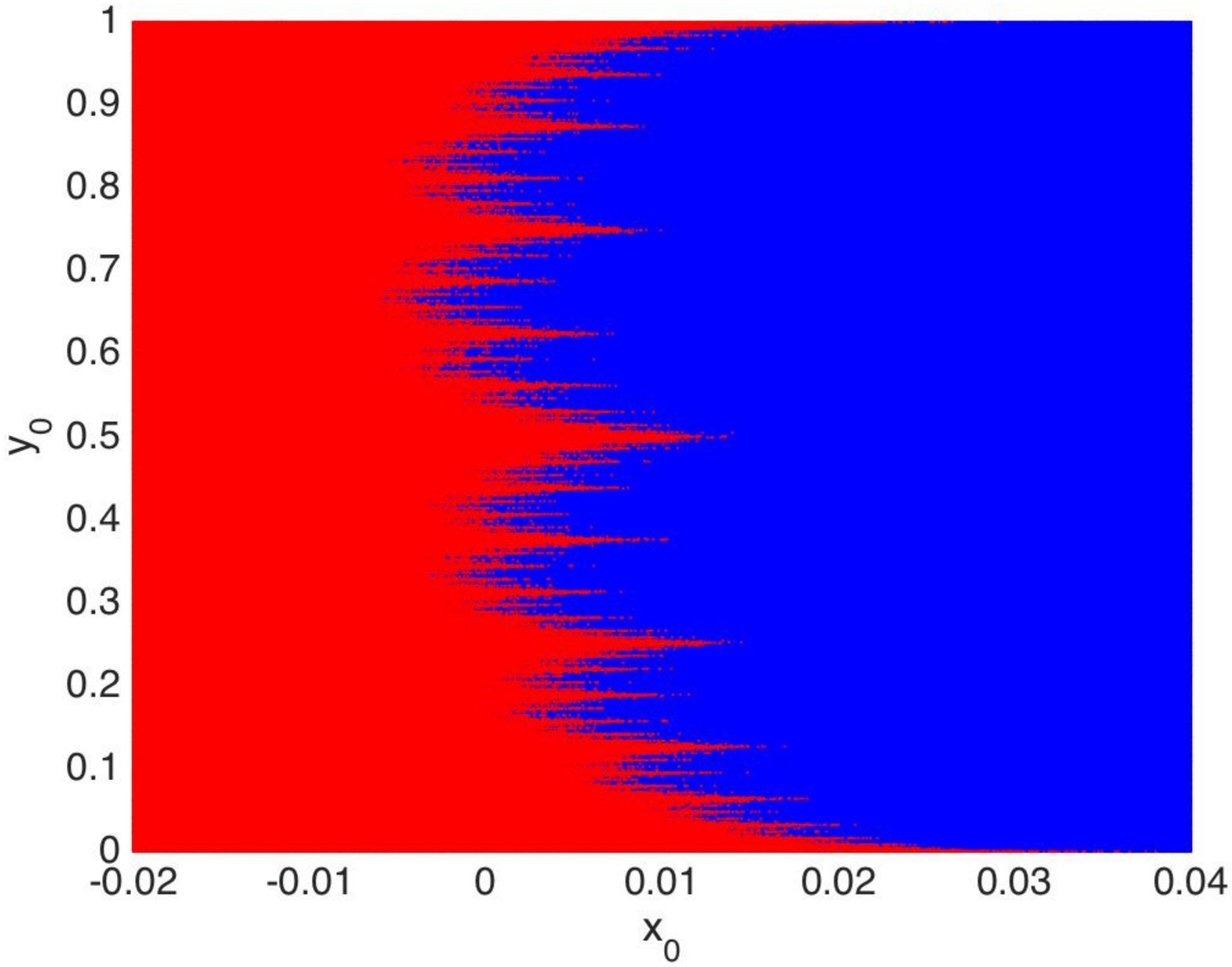}} 
        \caption{\label{fig:bb} 
        Basins of attraction in the system (\ref{eq:lin_1})-(\ref{eq:tent}), depicted by different colours/shades. $2^{22}$ randomly sprinkled initial conditions are used. }
    \end{center}
\end{figure*}

It can be visually checked in the case of our simple 2D model that the basin boundary is indeed not filamentary, but rather a jagged, rough curve. To this end we initialise many trajectories randomly sprinkled in a box containing the basin boundary, and color small markers placed in these initial positions differently with respect to the different attractor reached. The result of this is shown in Fig. \ref{fig:bb}. 

\begin{figure*} [t!]
    \begin{center}
            \scalebox{0.5}{\includegraphics{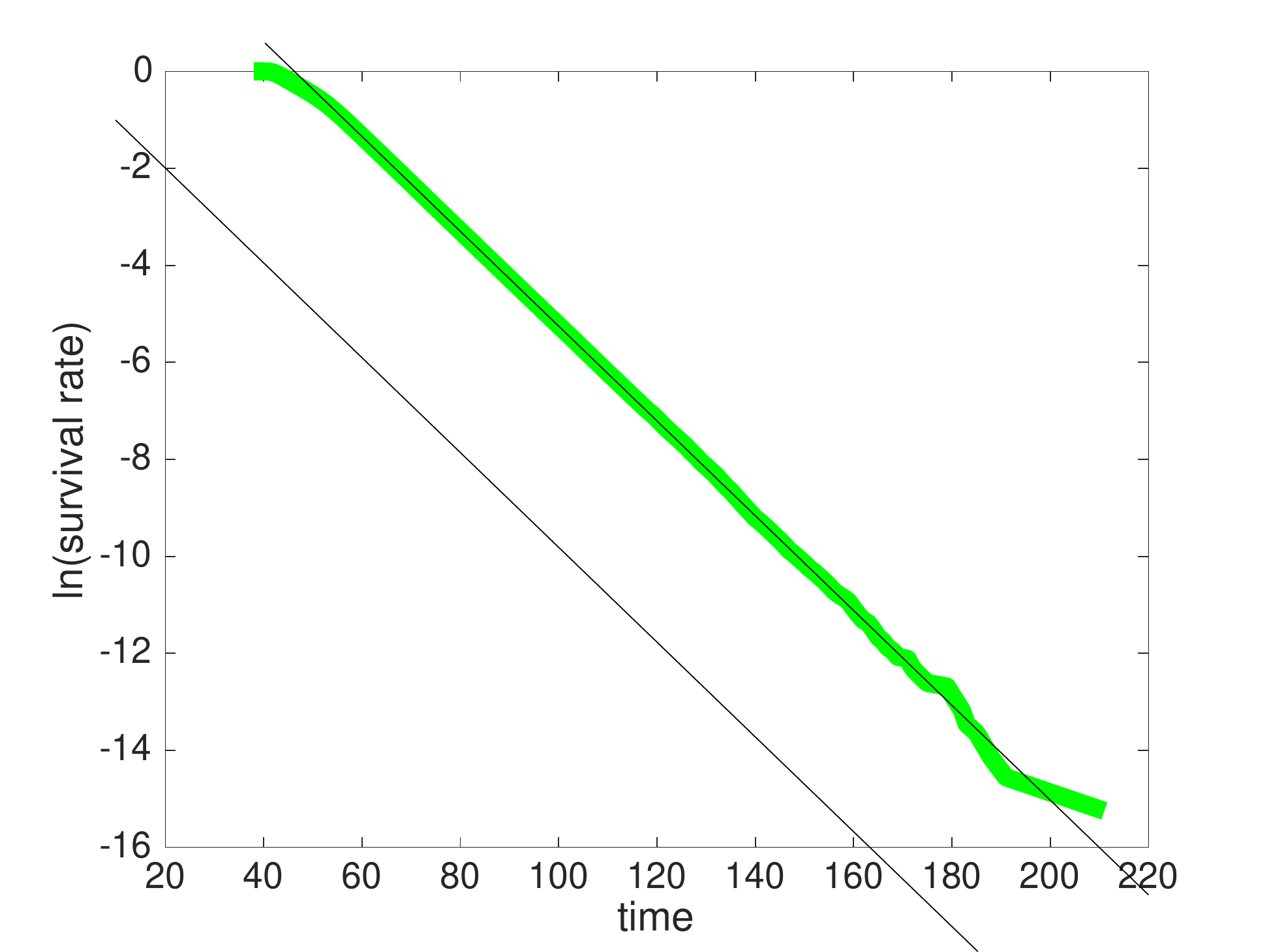}} 
        \caption{\label{fig:surv} 
        Survival rate based on the initial conditions seen in Fig. \ref{fig:bb}. The slope on the log-lin diagram gives the escape rate $\kappa$. Parallel straight black lines aid reading off $\kappa$.}
    \end{center}
\end{figure*}

\begin{figure*} %[t!]
    \begin{center}
        \begin{tabular}{cc}
            \includegraphics[width=0.5\linewidth]{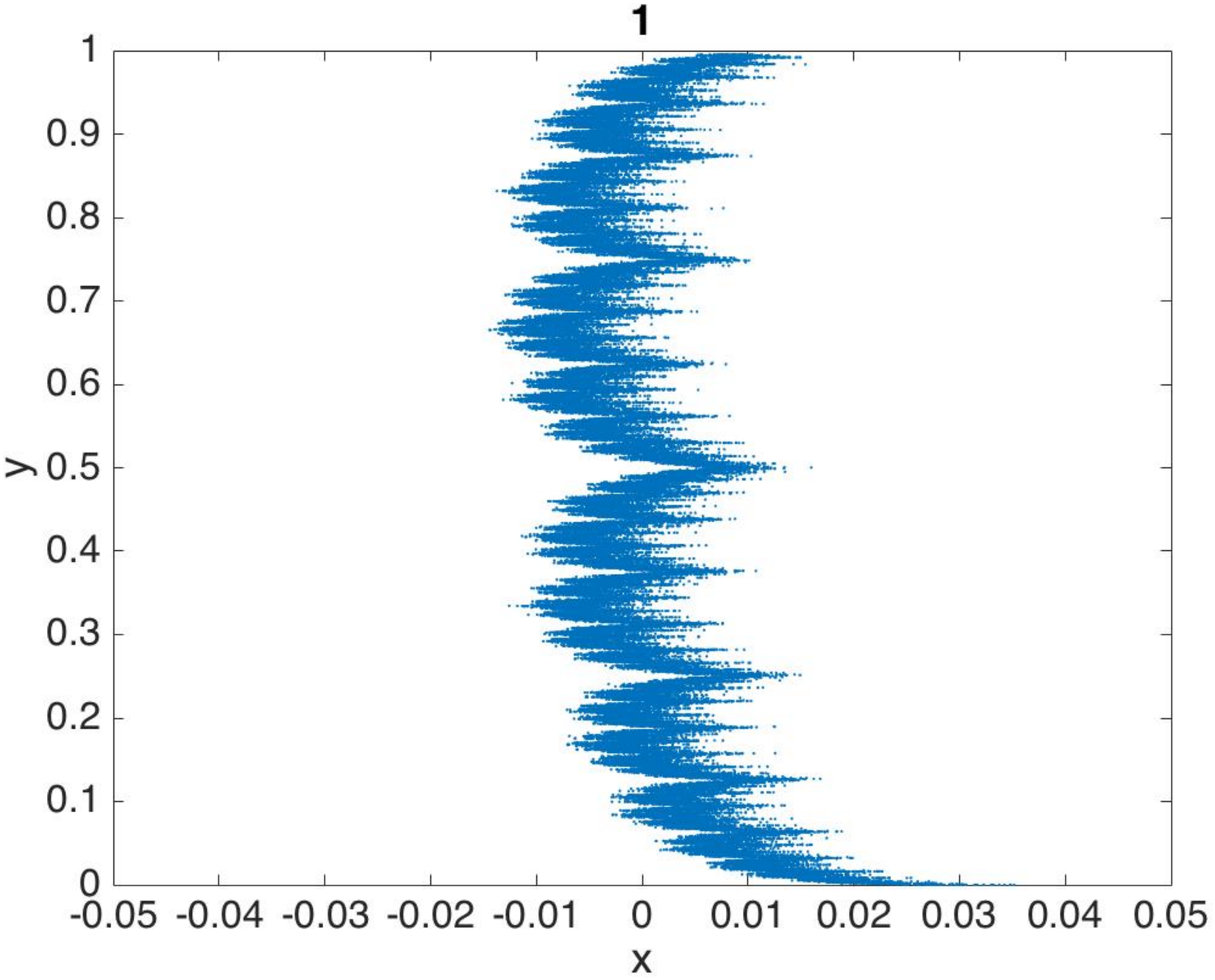} &
            \includegraphics[width=0.5\linewidth]{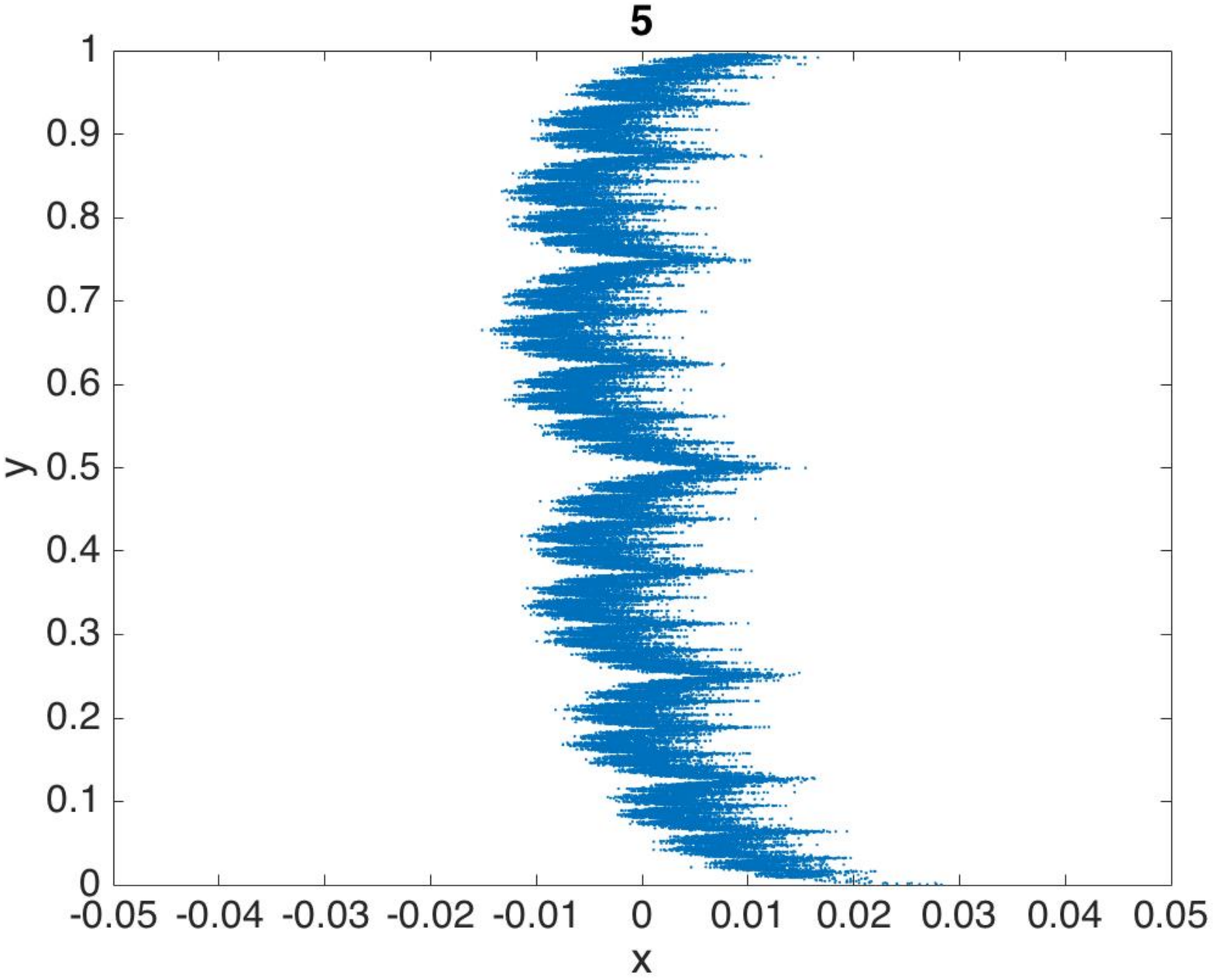} \\
            \includegraphics[width=0.5\linewidth]{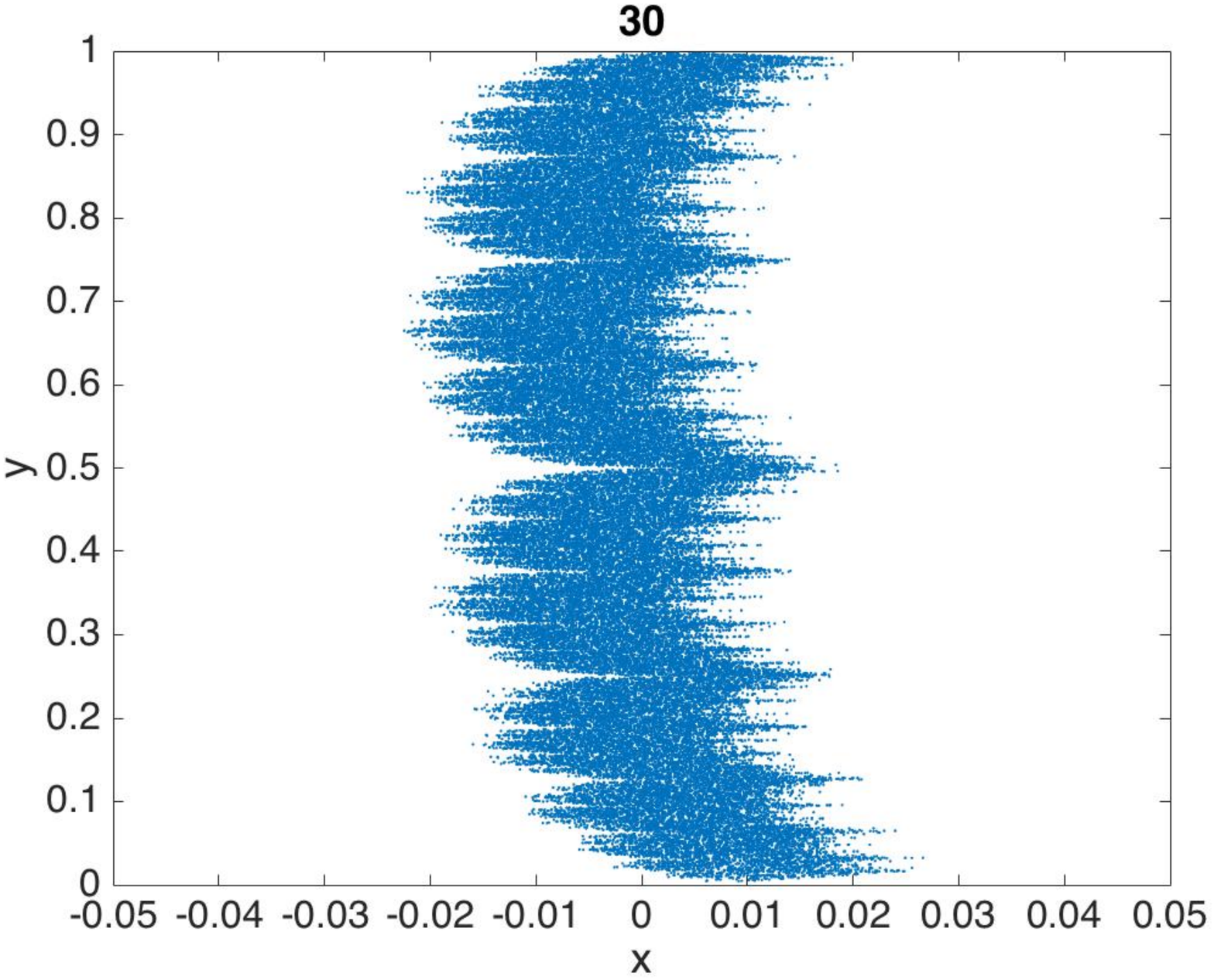} &
            \includegraphics[width=0.5\linewidth]{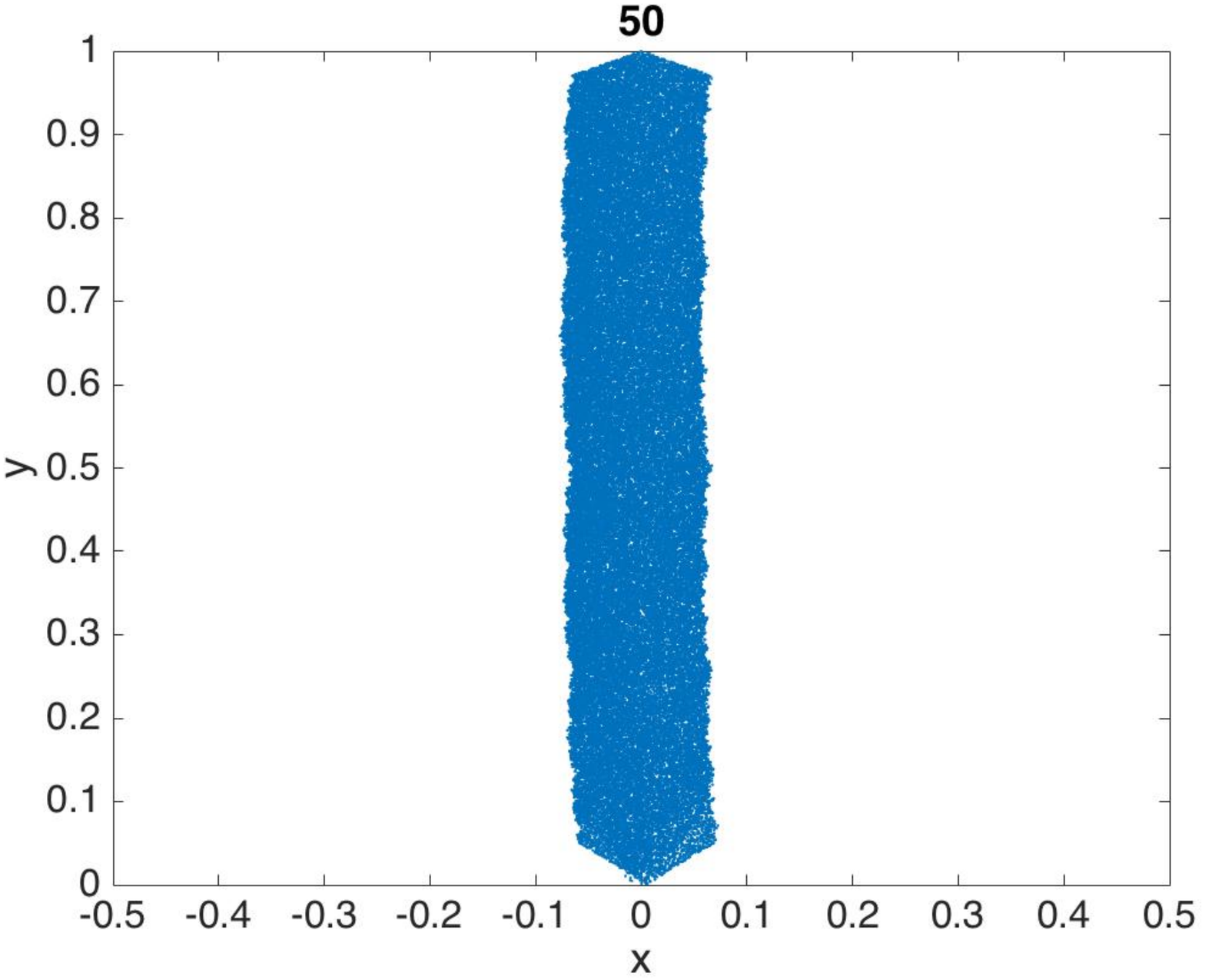} \\
            \includegraphics[width=0.5\linewidth]{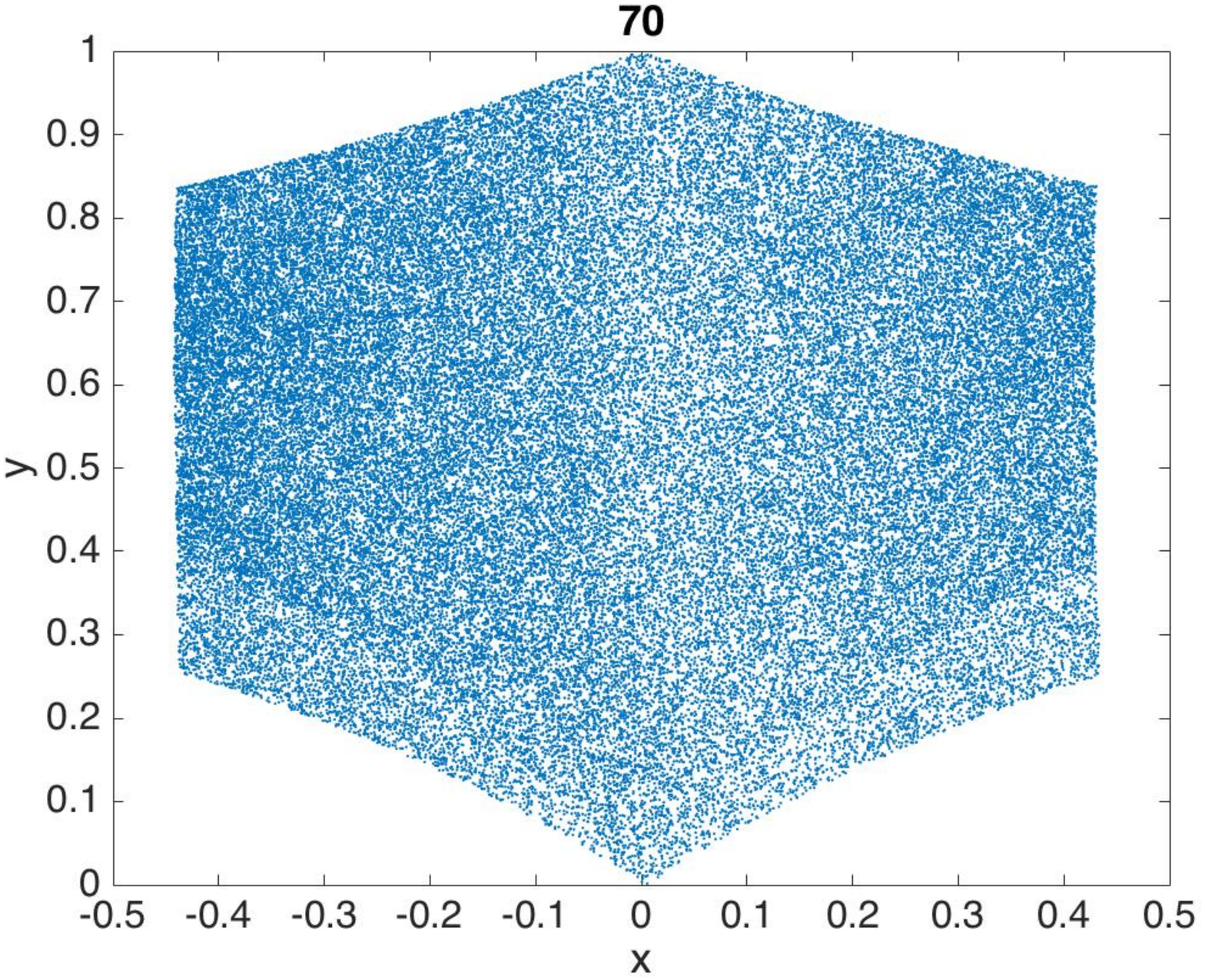} &
            \includegraphics[width=0.5\linewidth]{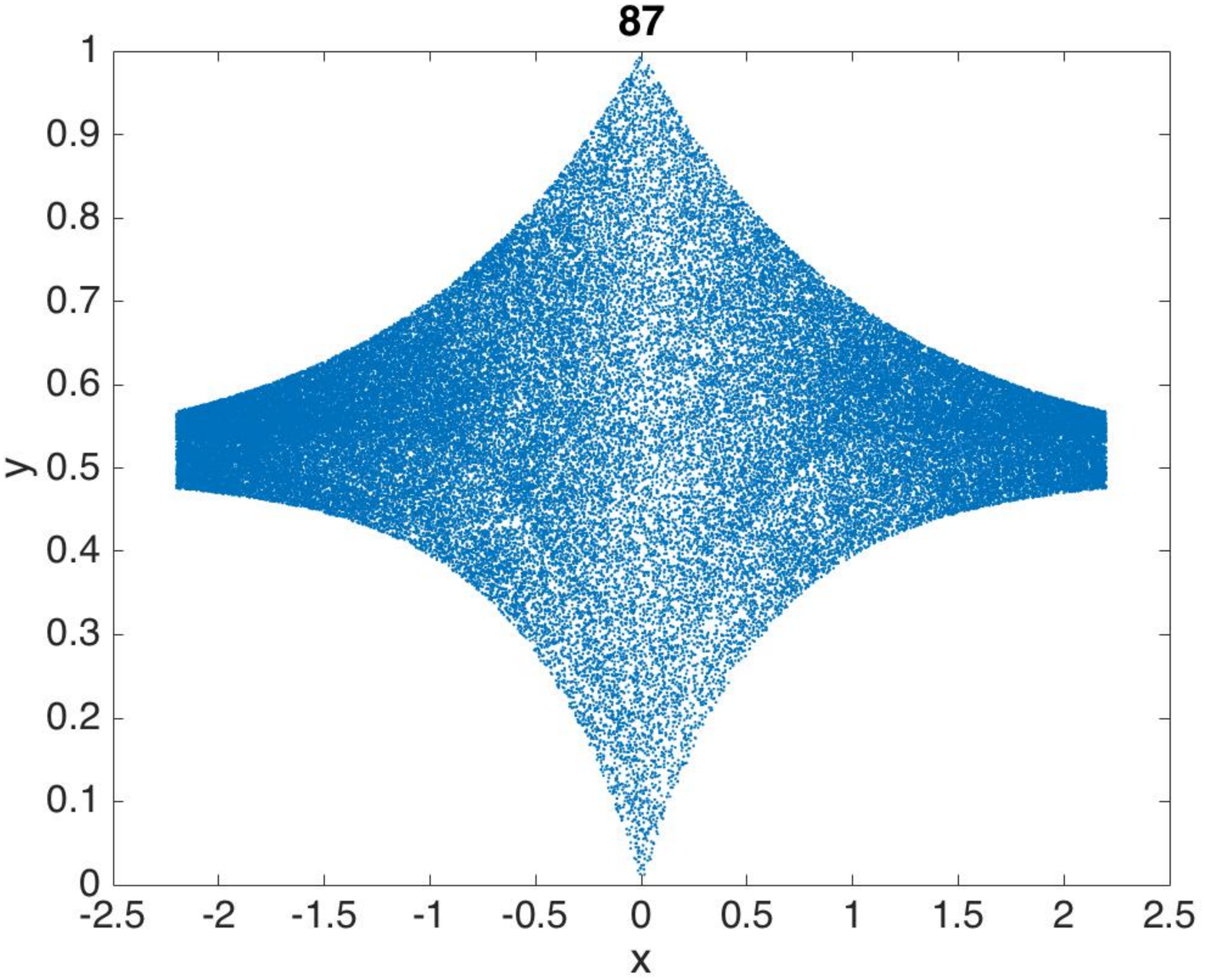}
        \end{tabular}
        \caption{\label{fig:sprinkler} Snapshots from the video following the ensemble used for the sprinkler method applied to the system (\ref{eq:lin_1})-(\ref{eq:tent}). Above each panel integer numbers show the iteration number/time passed.}
    \end{center}
\end{figure*}

This ensemble can be used to check the exponential decaying of the survival rate {\bfac or probability} in the box over time, shown in Fig. \ref{fig:surv}, by which we can select an exponentially small fraction of the ensemble on the initial configuration that give birth to long-lived trajectories. These initial conditions (IC) are very near the basin boundary, which is the stable manifold of the nonattracting set embedded in it. Therefore, the small fraction of long-lived trajectories should approach the nonattracting set as they evolve {\bfac (if it is different from the boundary itself---unlike in our case)}, and then leave the box along the unstable manifold of the nonatracting set. That is, with this procedure, called the sprinkler method~\cite{LT:2011}, one can construct the nonattracting set and its stable and unstable manifolds. Snapshots of the long-lived trajectory ensemble are displayed in Fig. \ref{fig:sprinkler}, revealing that the stable manifold is identical with the nonattracting set\footnote{It is interesting to note that despite the long time average $\langle y_i\rangle_i=0$, making the perturbation in (\ref{eq:lin_1}) symmetric around the unperturbed fixed point $x=0$, the average on the nonattrcating set $\langle x_i\rangle_i\neq0$.}, and that the unstable manifold is space filling. 
%Based on generic dimension formulae established in~\cite{PhysRevE.54.4819}, Sec. 8.3.1 of~\cite{LT:2011} shows that the dimension of the unstable manifold of the model in~\cite{GOY:1983} similar to ours (and in fact for any 2D maps with two positive LEs) is $D_{u,1}=2$, and that of the stable manifold is that of the nonattracting set itself, $D_{s,1}=D_1$, in agreement with our numerical finding.
{\bfac These were indeed the predictions derived from generic dimension formulae in Sec.~\ref{sec:dim_formulae} and in Sec. 8.3.1 of~\cite{LT:2011}.}
These properties are, again, very different from properties of nonattracting chaotic saddle sets embedded in {\em filamentary} fractal basin boundaries of dissipative invertible systems, which saddles have a signature double-fractal geometry, and so their stable and unstable manifolds both consist of filaments, which define two noninteger partial fractal dimensions; see Chapter 6 of~\cite{Tel_n_Gruiz:2006}.

\begin{figure*} [t!]
    \begin{center}
            \scalebox{0.5}{\includegraphics{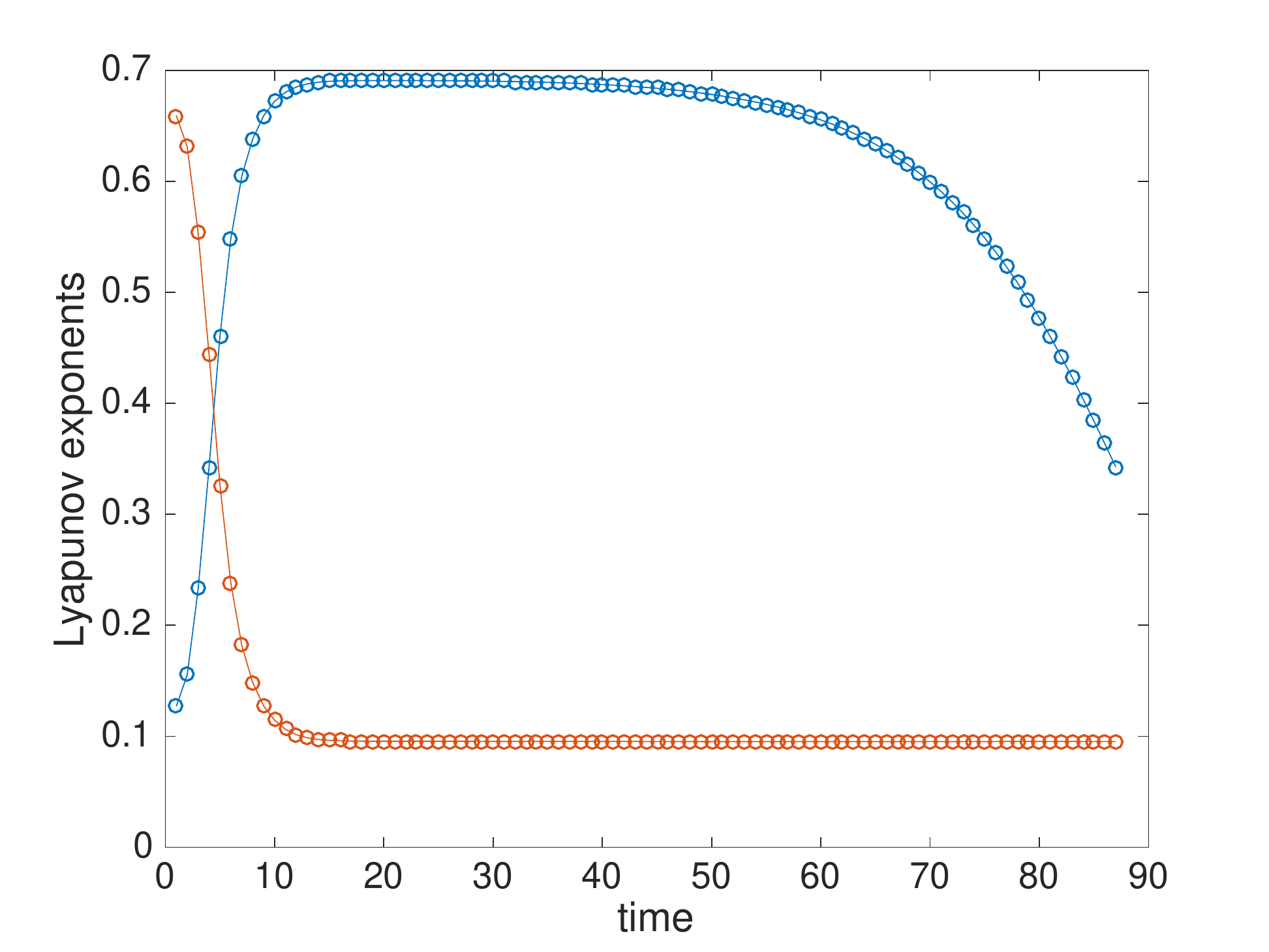}} 
        \caption{\label{fig:LEs} 
        Estimates of the two Lyapunov exponents on the nonattracting set of the system (\ref{eq:lin_1})-(\ref{eq:tent}). The actual estimates are those after a transient is past, at about iteration \#20. During the transient the Lyapunov (Gram-Schmidt) vector belonging to the blue line turns from a horizontal orientation to a vertical one. The discrete-time data points are linked by straight lines to guide the eye in reading the diagram.}
    \end{center}
\end{figure*}

We note that from the survival decay shown in Fig. \ref{fig:surv} we can extract the escape rate $\kappa$, which supports 
%that the result of ~\cite{SWEET20001} (\ref{eq:kappa}) applies to (\ref{eq:tent})-(\ref{eq:lin_1}). 
our claim that it is just the uncoupled/unperturbed $\tilde{\lambda}_x$ given weak perturbations and a regular unperturbed boundary, $\tilde{D}_1^{(x)}=0$. On the other hand, the numerically estimated Lyapunov exponents, shown in Fig. \ref{fig:LEs}, also agree with our claim under (\ref{eq:LEs}). %the formulae in~\cite{GOY:1983,SWEET20001} (\ref{eq:LEs}). 
We estimated the LEs using the Gram-Schmidt procedure. Since the LEs of interest belong to the nonattracting set, we first have to construct that set {\bfac and its natural measure}. It has in fact been already done using the sprinkler method (Fig.~\ref{fig:sprinkler}). The estimates of the LEs at any time are taken as averages of one-step LEs over the ensemble of long-lived trajectories followed.

The analytical formulae for the Lyapunov exponents given in~\cite{GOY:1983} for a system similar to (\ref{eq:lin_1})-(\ref{eq:tent}) are identical to (\ref{eq:LEs}). An analytical formula for the escape rate given in~\cite{SWEET20001} for a generalised model of that in~\cite{GOY:1983} implies for the special case of~\cite{GOY:1983} that the escape rate equals the cross-boundary LE; this is in agreement again with our claim based on our heuristic argument. In~\cite{KM-PY:1984} the authors derive a formula also for the fractal dimension $D_{b,0}$ which is identical to what we derive from %the 
generic relations %(\ref{eq:uncert}) and (\ref{eq:kappa_sum}) 
for the case of weak perturbations as (\ref{eq:part_dim_x}) and (\ref{eq:dim_pred_reg}).

% \begin{equation} \label{eq:KM-PY84_1}
%  D_{b,0}=2-\lambda_u/\lambda_p=2-\ln a/\ln2.
% \end{equation} 

In Sec. \ref{sec:dim_numerics} we check if the said dimension formula applies indeed to our system. We will do this in the {\em invertible} 3D ($D=3$) system:

% \begin{eqnarray}
%  x_{n+1} &=& ax_n + d(z_n - 1/2)  \label{eq:lin_2} \\
%  y_{n+1} &=& 
%  \left\{ 
%   \begin{tabular}{ll}
%   $cy_n$       & $z_n>1/2$ \\
%   $1+c(y_n-1)$ & $z_n\leq1/2$ 
%   \end{tabular}
%  \right. \label{eq:baker_1} \\
%  z_{n+1} &=& 
%  \left\{ 
%   \begin{tabular}{ll}
%   mod$(2z_n+x_n,1)$       & $z_n>1/2$ \\
%   mod$(1+2(z_n-1)+x_n,1)$ & $z_n\leq1/2$ 
%   \end{tabular}
%  \right. \label{eq:baker_2}
% \end{eqnarray} 
% Swap y, z, so that \lambda_y is the positive LE also here! Btw. it was that way earlier, bb_global_instab_02.pdf, and i cannot remember why i swapped that.
\begin{eqnarray}
 x_{n+1} &=& ax_n + d(z_n - 1/2)  \label{eq:lin_2} \\
 y_{n+1} &=& 
 \left\{ 
  \begin{tabular}{ll}
  mod$(2y_n+x_n,1)$       & $y_n>1/2$ \\
  mod$(1+2(y_n-1)+x_n,1)$ & $y_n\leq1/2$ 
  \end{tabular}
 \right. \label{eq:baker_2} \\
 z_{n+1} &=& 
 \left\{ 
  \begin{tabular}{ll}
  $cz_n$       & $y_n>1/2$ \\
  $1+c(z_n-1)$ & $y_n\leq1/2$ 
  \end{tabular}
 \right. \label{eq:baker_1}
\end{eqnarray} 
where the linear equation (\ref{eq:lin_2}) (identical with eq. (\ref{eq:lin_1})) is perturbed by the invertible chaotic baker map (\ref{eq:baker_2})-(\ref{eq:baker_1}). The behaviour wrt. the basin boundary is similar (results not shown) as in the 2D noninvertible system (\ref{eq:lin_1})-(\ref{eq:tent}). % for a reason that will be clarified later. ***WILL IT BE? OR ALREADY OBVIOUS?
We will use $c=1/3$. However, note that the fractal dimension of the boundary does not depend on $c$; {\bfac see eq. (\ref{eq:uncert})}. This 3D model is a minimal model for an invertible system featuring a rough basin boundary with a dissipative dynamics on the nonattracting chaotic set. The nonattracting set in the invertible system can therefore be called a saddle. In contrast, the dynamics on the nonattracting chaotic set of the 2D minimal model (\ref{eq:lin_1})-(\ref{eq:tent}), called a repellor, is not dissipative.

\subsection{Mixed filamentary and rough boundary}

\begin{figure*} %[t!]
    \begin{center}
        \begin{tabular}{cc}
            \includegraphics[width=0.5\linewidth]{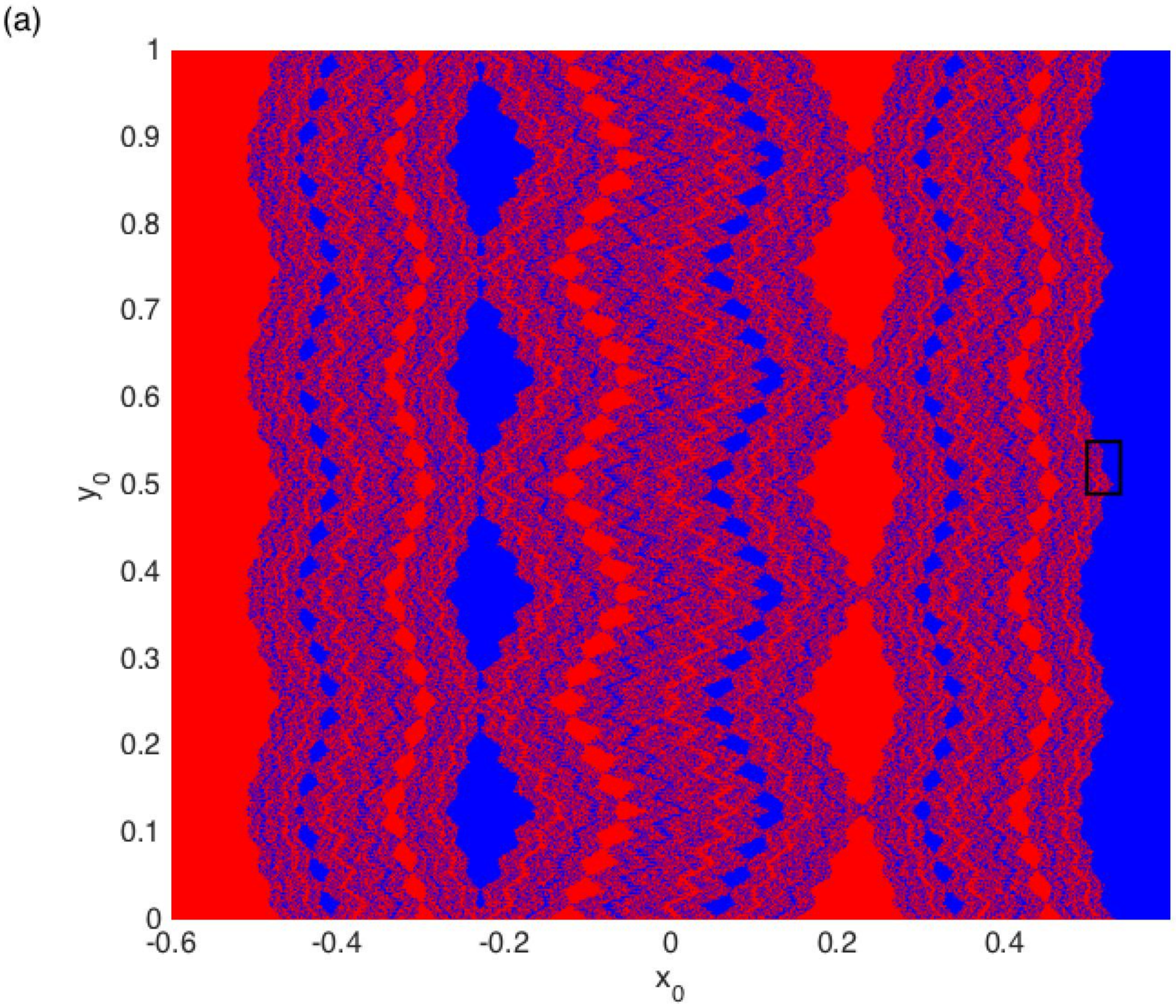} &
            \includegraphics[width=0.5\linewidth]{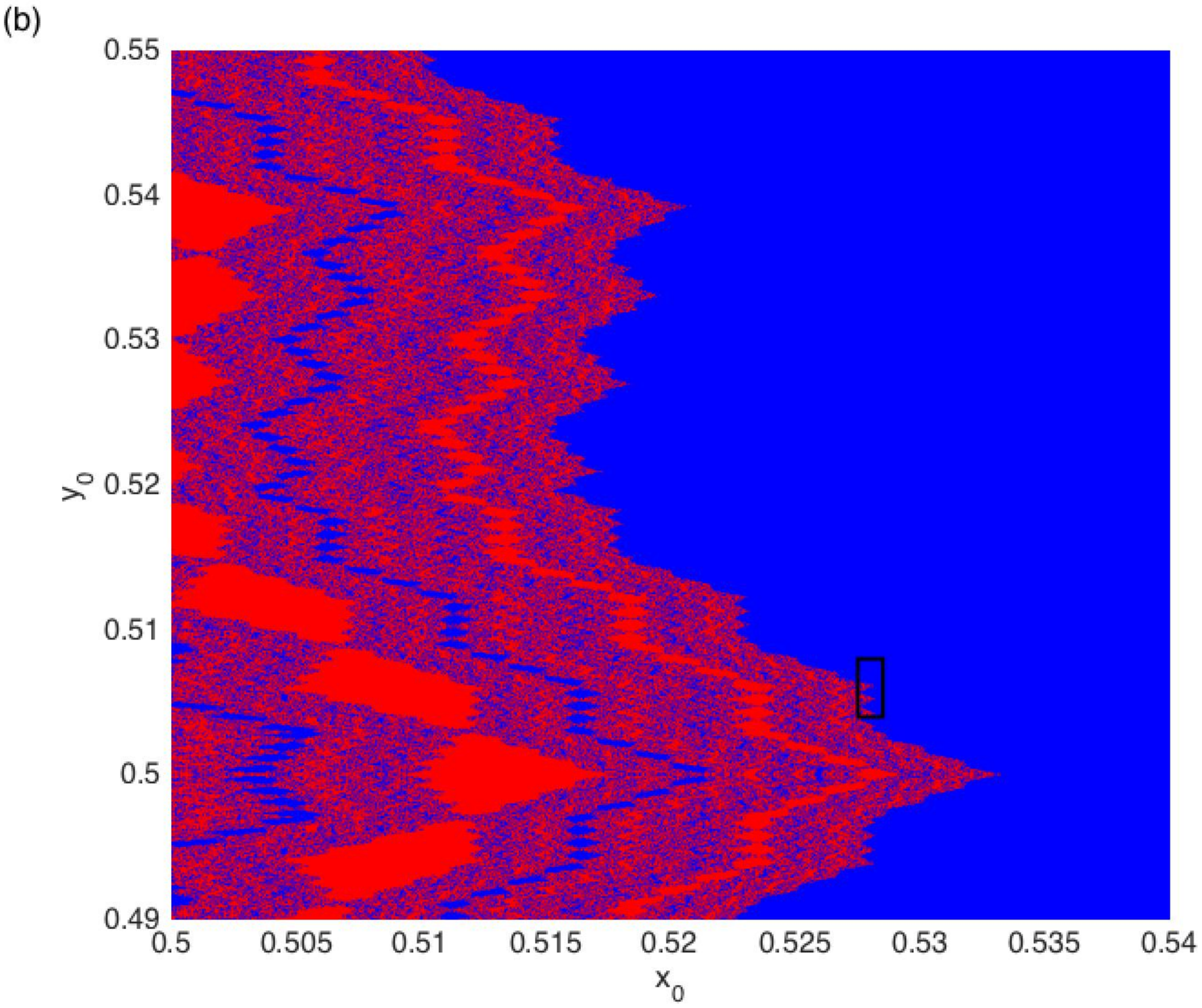} \\
            \includegraphics[width=0.5\linewidth]{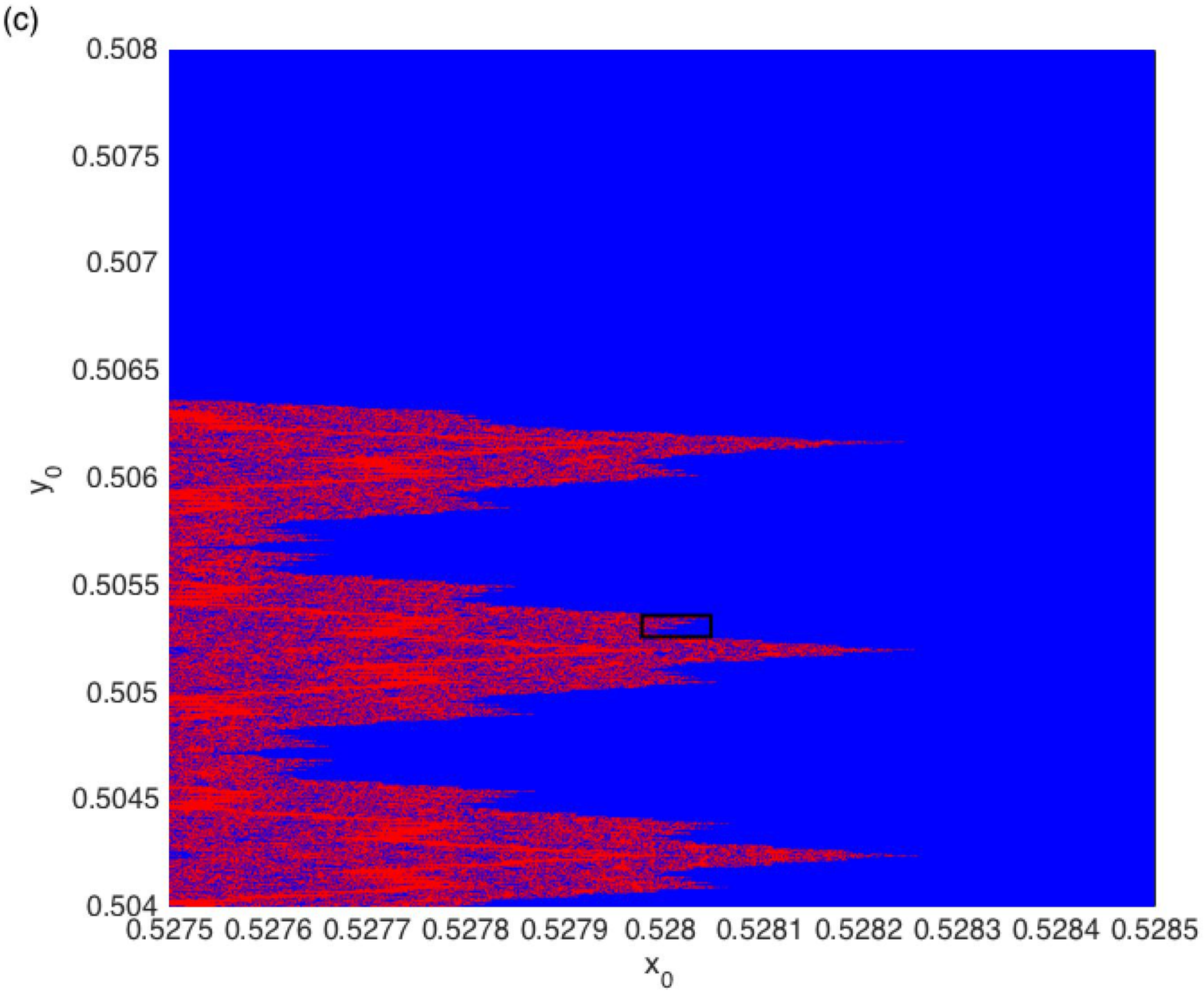} &
            \includegraphics[width=0.5\linewidth]{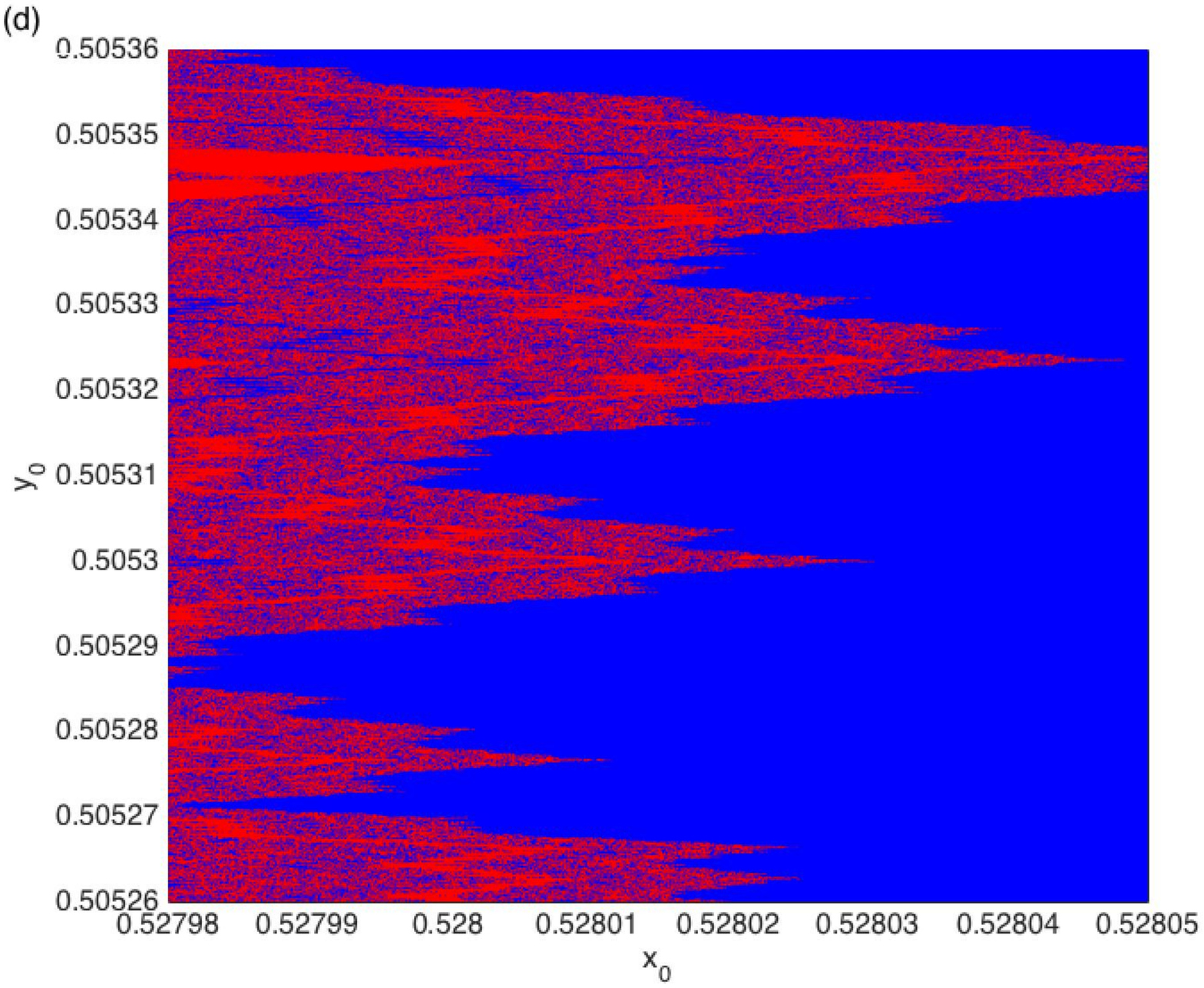} 
        \end{tabular}
        \caption{\label{fig:mixed_rough_filamentary} Visual display of a new type of rough basin boundary via the basins of attraction. As in Fig. \ref{fig:bb}, ICs colored in red (blue) go to the left (right). (a) shows the entirety of the boundary and (b), (c), (d) provide three successive zooms onto features of the rightmost ``filament''. The zoomed areas are indicated by black frames.}
    \end{center}
\end{figure*}

To have a filamentary fractal unperturbed boundary (or perturbed but not rough), we replace the linear eq. (\ref{eq:lin_1}) of the cross-boundary dynamics by a nonlinear one similar to that studied in~\cite{PGL:1992}:   

\begin{eqnarray}
 x_{n+1} &=& d(y_n-1/2) + 
 \left\{ 
  \begin{tabular}{ll}
  $a(x_n+1/2)-1/2,$       & $x_n<\gamma,$ \\
  $-bx_n,$ & $|x_n|<\gamma,$  \\
  $a(x_n-1/2)+1/2,$       & $x_n>-\gamma,$
  \end{tabular}
 \right. \label{eq:mixed_2} \\
 y_{n+1} &=& \mu/\beta\min[\mod(y_n,\beta),\beta-\mod(y_n,\beta)],\ \mu = 1 + \exp(c|x_n|), \label{eq:mixed_1} 
\end{eqnarray} 
where $\beta=2^{1-q}$, $\gamma=(1-a)/(a+b)/2$, $a>2$ and $ b>a/(a-2)$. Following the methodology of Sec. 2.2.3 of~\cite{LT:2011} applied to a similar but not bistable model, we can derive analytical formulae for the unpertubed LEs:

\begin{equation}\label{eq:mixed_LEs}
 \tilde{\lambda}_x = \frac{2b\ln a + a\ln b}{2b+a}, \quad \tilde{\lambda}_y = q\ln 2,
\end{equation} 
and a formula also for the unperturbed escape rate:

\begin{equation}
 \tilde{\kappa} = \ln\left(\frac{ab}{2b+a}\right).
\end{equation} 
Of these we need $\tilde{\kappa}$ and $\tilde{\lambda}_y$ to predict the co-dimension of a rough boundary, on the one hand, by eq. (\ref{eq:part_dim_y}) as:

\begin{equation} \label{eq:dim_mixed}
 D_1^{(y)} \approx 1- \frac{\ln\left(\frac{ab}{2b+a}\right)}{q\ln 2}.
\end{equation} 
On the other hand, by predicting $\lambda_x$ and $\lambda_y$, we can predict for what parameter choices do we actually get a rough boundary. Note that for $q=1$ eq. (\ref{eq:mixed_1}) is that of the tent map (\ref{eq:tent}). However, with that choice no valid choices of $a$ and $b$ exist to yield a rough boundary, which is the very reason why we use the ``camping'' (or ``multi-tent'') map (\ref{eq:mixed_1}) instead. For reliable numerical verification of the dimension formula (\ref{eq:dim_mixed}) we need as large a contribution $\hat{D}_1$ from roughness, given by (\ref{eq:dim_rough}), as possible. Reasonably favorable parameter choices for this objective we have found as: $a=4$, $b = \pi/6 + a/(a-2)\approx 2.52$, $c=-1$, $d=0.2$, $q=3$, for which we predict $\lambda_x=1.18$, $\lambda_y=2.08$, $\kappa=0.11$, $D_1^{(y)}=0.947$, $\hat{D}_1=0.04$. In Fig. \ref{fig:mixed_rough_filamentary} we visualize the boundary, providing successive zooms on the rightmost ``filament'' in order to expose its roughness. The provided value for $\kappa$ has been checked to apply---to the given approximation---to both the unperturbed and perturbed dynamics, indicating that $d$ is small enough. Yet, it is large enough to make the ``filaments'' not only rough but seemingly intertwined.

\section{Numerical calculation of the fractal dimension}\label{sec:dim_numerics}

The co-dimension $D-D_{b,0}$ of the basin boundary is maximum 1, as it has to act as a separator of different regimes of phase space. Because of this, it is %can be thought to be 
possible to determine this co-dimension as the co-dimension $1-D_0^{(c)}$ of the {\em intersection set} of the boundary and a straight line traversing it. This technique is easily applicable in any high-dimensional system, as demonstrated in~\cite{0951-7715-30-7-R32}.  
%However, is it really so that $D-D_{b,0}=1-D_0^{(c)}$; or is it rather $D_0^{(c)}=D_0^{(x)} = 1$, according to eq. (\ref{eq:part_dim_x}) {\bfa and Fig. \ref{fig:under_det_dim}}? We demonstrate here that it is the former. 
{\bfb It is important to appreciate that the formula (\ref{eq:dim_intersect}) is satisfied for {\em almost any} choice of the angle that the traversing line (set $S_1$) makes with the boundary (set $S_2$), i.e., with probability one for a random choice. Therefore, even if we found that the cross-boundary partial dimension is trivially $D_0^{(x)} = 1$, applying the said simple method, we will actually measure the nontrivial co-dimension. Therefore, even if we wanted to utilize this distinguishing characteristic of a rough boundary for a detection purpose, we cannot.}

\begin{figure*} [t!]
    \begin{center}
            \scalebox{0.5}{\includegraphics{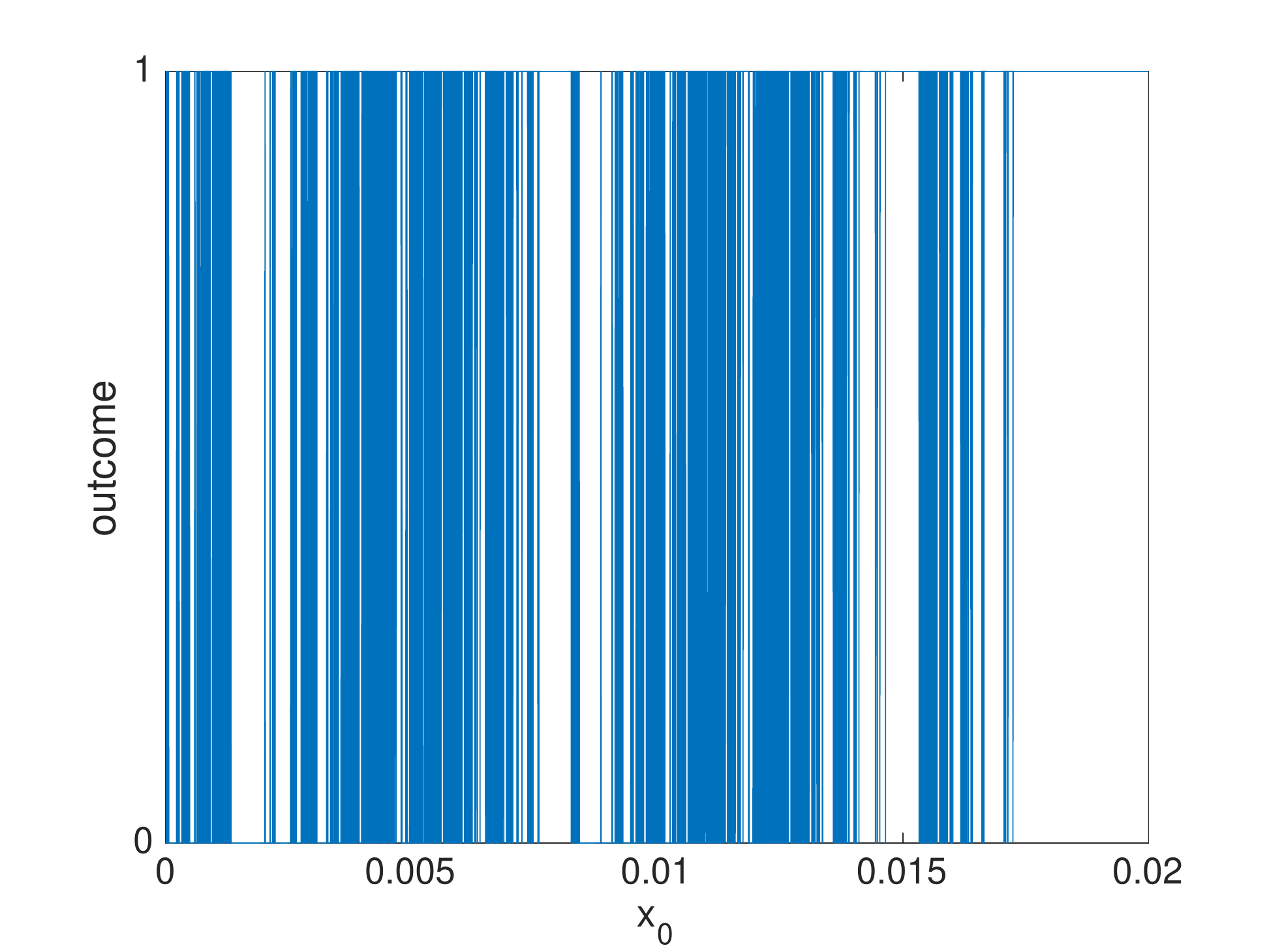}} 
        \caption{\label{fig:06} 
        The trajectory of (\ref{eq:lin_2})-(\ref{eq:baker_1}) escapes to positive (negative) infinity wrt. $x$ when the outcome is 1 (0). $2^{17}$ equally spaced ICs of $x_0$ are examined in the shown range, while $y_0$, $z_0$ are fixed at some arbitrary values.}
    \end{center}
\end{figure*}

\begin{figure*} [t!]
    \begin{center}
            \scalebox{0.5}{\includegraphics{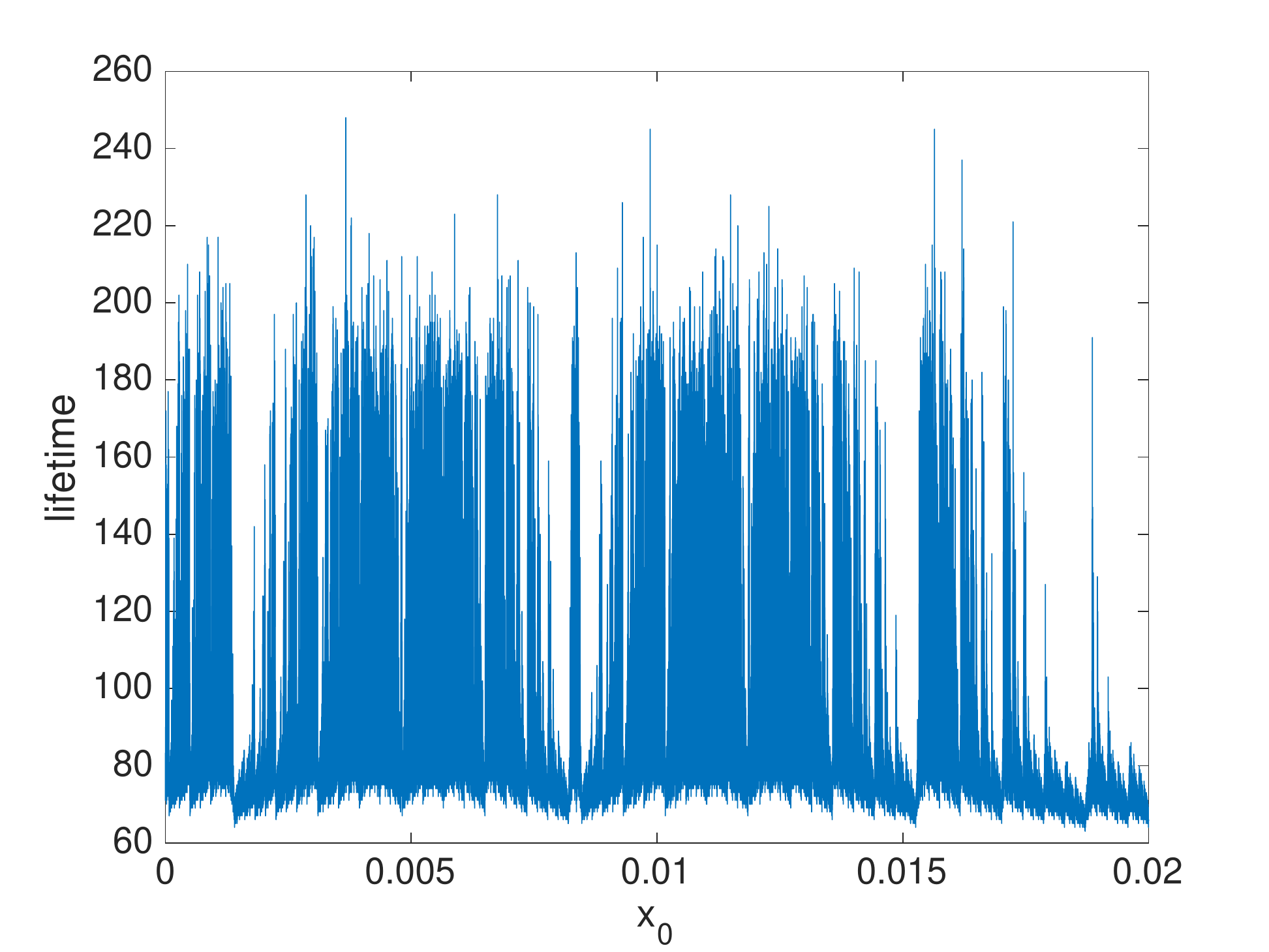}} 
        \caption{\label{fig:01} 
        Trajectory lifetimes in (\ref{eq:lin_2})-(\ref{eq:baker_1}) belonging to initial conditions $x_0$ (and $y_0$, $z_0$ fixed at some arbitrary values). $2^{25}$ equally spaced ICs of $x_0$ are examined in the shown range.}
    \end{center}
\end{figure*}

On the technical side, it is not immediately clear how to obtain sample points of this intersection set. Instead, the co-dimension can supposedly be determined, being equal to the uncertainty exponent $\alpha$, based on data generated as follows. We populate a $y,z=constant$ line by equispaced ICs, and let them evolve under (\ref{eq:lin_2})-(\ref{eq:baker_1}) until they approach one of the attractors, determining thereby the outcome. A visual representation of the outcomes along the line is given in Fig. \ref{fig:06}. It is accompanied by a complicated and seemingly discontinuous function of the lifetimes of the trajectories depending on the ICs, shown in Fig. \ref{fig:01}.

\begin{figure*} [t!]
    \begin{center}
            \scalebox{0.5}{\includegraphics{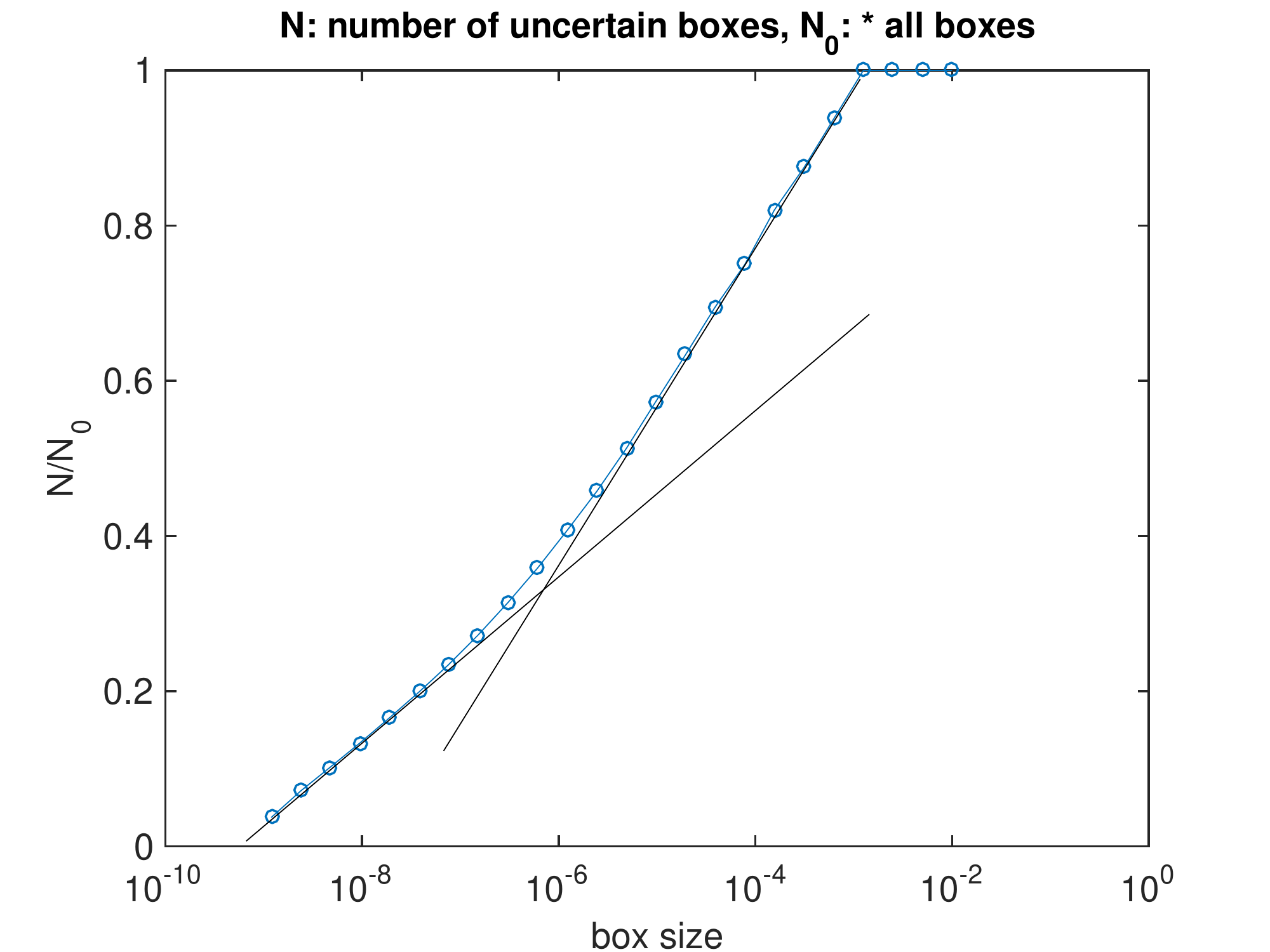}} 
        \caption{\label{fig:02} 
        Number of uncertain boxes derived from the initial conditions examined in Fig. \ref{fig:01} versus the linear box size $\epsilon$ in a log-lin diagram. Straight black lines indicate regimes. The discrete data points are linked by straight lines to guide the eye in reading the diagram.}
    \end{center}
\end{figure*}

\begin{figure*} [t!]
    \begin{center}
            \scalebox{0.5}{\includegraphics{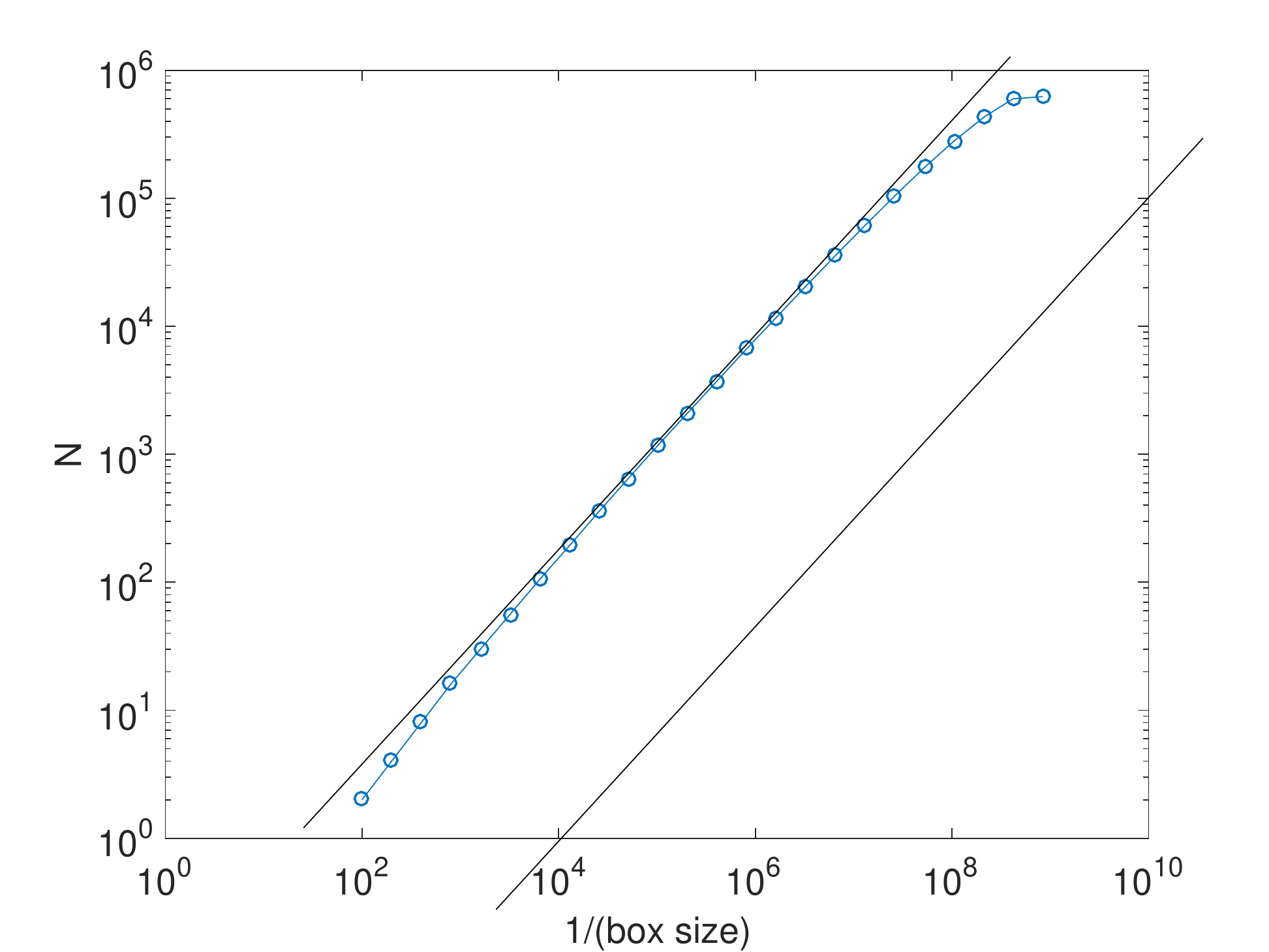}} 
        \caption{\label{fig:03} 
        Number of uncertain boxes derived from the initial conditions examined in Fig. \ref{fig:01} versus the linear box size $\epsilon$ in a log-log diagram. Parallel straight black lines indicate the steepness of the slope.}
    \end{center}
\end{figure*}

The ratio $N/N_0$ of uncertain boxes, in which we have different outcomes, of linear size $\epsilon$ is shown in Fig. \ref{fig:02}. Values of $N/N_0$ are plotted on a logarithmic scale, and those of $\epsilon$ are on a linear scale. In the diagram two regimes can be seen, in both of which the decay is exponentially fast, only the scale (the slope in the log-lin diagram) is different. That is, the decay is faster than the usual polynomial decay that defines the uncertainty exponent~\cite{LT:2011}: $N/N_0 \sim \epsilon^{\alpha}$. Indeed, plotting in a log-log diagram, seen in Fig. \ref{fig:03}, the data does not show a scaling of good quality in any range; the line is curved. {\bfb The significance of this experience as a negative result is that} this alone (without the theory presented in Sec. \ref{sec:condition}) could call into question whether the basin boundary has a fractal geometry {\bfc as we know it} in our case. {\bfb Next, we proceed with the alternative method by which we could actually verify the fractality of the set and our prediction for the dimension.}

\begin{figure*} [t!]
    \begin{center}
            \scalebox{0.5}{\includegraphics{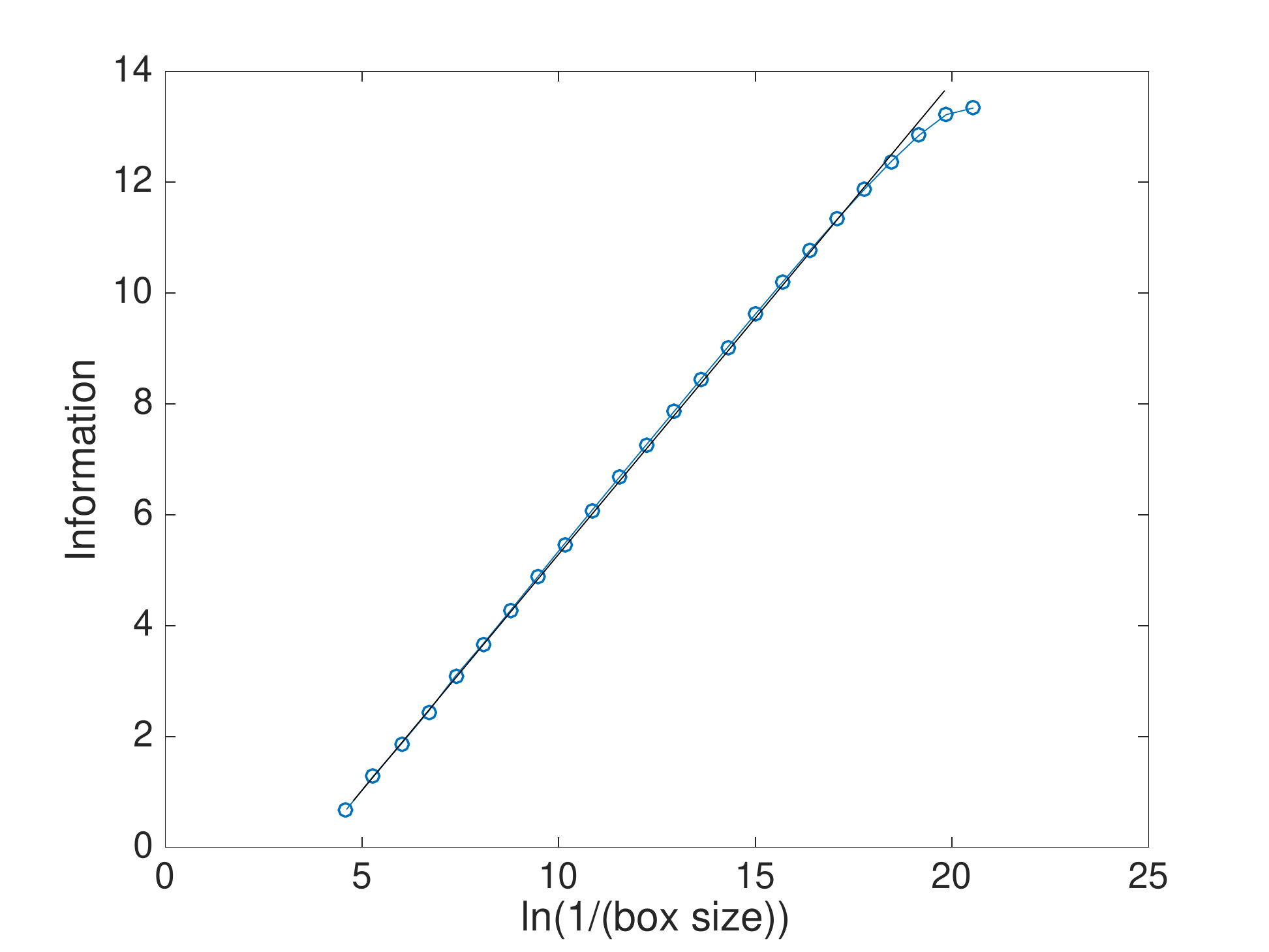}} 
        \caption{\label{fig:04} 
        The information derived from the initial conditions examined in Fig. \ref{fig:01} versus the linear box size $\epsilon$. A straight black line indicate the quality of the scaling. The slope of the fitted straight line by least-squares is 0.84, and the room-mean-square-error of the fit is 0.04. %D_1 = 0.84165 , RMSE = 0.043175
        }
    \end{center}
\end{figure*}

One can apply `formally' a box-counting algorithm to the outcome data to estimate the {\em information} dimension $D_1^{(c)}$, as if the points where the outcome is reversed by a small shift of the IC corresponded to trajectory data points. The information {\bfac $I(\epsilon)=-\sum P_i(\epsilon)\ln P_i(\epsilon)$, $P_i$ being normalised box counts~\cite{Tel_n_Gruiz:2006},} as a function of the box size, or rather $-\ln(\epsilon)$, is shown in Fig. \ref{fig:04}. This features a scaling of good quality, {\bfb which, first, verifies fractality (beyond visuals), and, second,} we estimate that $D_1^{(c)} = 0.84$. However, this estimate of the information dimension is appropriate only if the measure across the line over the supporting set is a Lebesgue measure, i.e., constant, because the outcome data does not contain information on the dynamics, only the geometry of the nonattracting set. 
We have already seen evidence in Fig. \ref{fig:sprinkler} that the measure is constant, namely, that the ensemble does not change with respect to its uniform distribution initially, achieved by random sprinkling; the distribution remains uniform. This backs the intuition that the constant measure of the uncoupled chaotic dynamics is inherited by the coupled dynamics. 
Therefore, $D_1^{(c)}=D_0^{(c)}$. 
This agrees with the authors of~\cite{SWEET20001} whose ``numerical computation indicates that for the repeller, the box-counting dimension $D_0$ and the information dimension $D_1$ are equal'', as also noted in Sec. 8.3.1.2 of~\cite{LT:2011}.
This fact has a relevance to the verification of dimension formulae given in Sec. \ref{sec:theory}, where $D_1$ appears instead of $D_0$, and if the measure is not uniform, with the above technique we can determine only $D_0$. What matters here is that 
%implies $D_{b,0} = 1.84$, which is in excellent agreement with the prediction of eqs. (\ref{eq:KM-PY84_1}) being $D_{b,0} = 1.86$.
%$D_0^{(c)} = 0.84$, that is, $D_0^{(c)}\neq D_0^{(x)}=1$ but % 11.01.2020.  V's contribution
{\bfb we measure $D_0^{(c)} = D_0^{(y)} = 0.84$, %on the one hand. And, on the other hand, the 
and this} numeric value agrees very well with our prediction by eq. (\ref{eq:dim_pred_reg}) being $D_0^{(y)}=0.86$.

Using the same method we have computed the partial dimension $D_0^{(y)}$ of the new type of rough basin boundary featured by (\ref{eq:mixed_2})-(\ref{eq:mixed_1}), too. Our result of 0.943 with a 0.007 root-mean-square-error of fitting a straight line to data is in rather good agreement with our prediction of $D_1^{(y)}=0.947$, even when considering that the contribution from roughness was predicted to be $\hat{D}_1=0.04$. Note that the roughness can be resolved only at sufficiently small length scales due to the weakness of the perturbation, as prompted also by Fig. \ref{fig:mixed_rough_filamentary}.

\section{Discussion and outlook}\label{sec:discuss}

{\bfc This work has been motivated by the finding of a fractal basin boundary in Ref.~\cite{0951-7715-30-7-R32} in a model whose bistable ocean component is not chaotic. There we conjectured that the condition for rough fractals established by Grebogi et al.~\cite{GOY:1983} in a 2D noninvertible map should apply. Here we show---following the approach of Vollmer et al.~\cite{VSE:2009}---that it is really the case. Furthermore, we derived a formula (\ref{eq:uncert}) for the co-dimension of this fractal basin boundary, which is a generalization of the Kantz-Grassberger (KG) formula holding only for filamentary fractal boundaries. We also found that this new formula implies unexpected properties of rough fractals, contrasting those of filamentary fractal boundaries, including the new three-way relationship of the predictability of the second kind and local and global instability. We continue discussing these now.}

We speculate that the formal result of $D_1^{(x)}=1$, {\bfac which is not excluded by us to hold also in higher dimensions}, owes perhaps to the fact that the boundary as a Weierstrass function is nondifferentiable and continuous in the same time. However, given that it is a function $W(y)$, the dimension $D_1^{(c)}$ of the intersection set of the boundary and a straight line $y,z=constant$ is actually 0. We have checked this numerically with $c=0$ in (\ref{eq:lin_1})-(\ref{eq:tent}). With two-way coupling, e.g. $c=-1$, we immediately have a $D_1^{(c)}>0$. 

{\bfb In the 2D invertible map that we analyzed in detail,} %{\bfa $x$ and $y$ meant to denote cross- and in-boundary coordinates, respectively,} % 11.01.2020. already in the Theory sec.
%formally $D_1^{(x)}=1$, which -- we speculate -- is a manifestation of the nondifferentiability of the boundary. 
it is, instead, $D_1^{(y)}<1$ that is not full-dimensional, suggesting that escape occurs not cross-boundary, but along the $y$-direction. %\footnote{{\bf DELETE}Note that in bistable systems escape occurs along a {\em single} direction, and the associated nonattracting chaotic set is called {\em low-dimensional}~\cite{LT:2011}. That is, the dimensionality of the nonattracting set in this sense is not to be confused with the dimensionality of the phase space or system.}. 
The intuitive picture that can be attached to this is that a trajectory has to be ``pushed'' in the $y$-direction into the ``crevices'' of the rough boundary ``landscape'' (into the gaps of the Cantor set whose fractal dimension is $D_1^{(y)}$) to be able to escape eventually in the $x$-direction. Although the local directionality of the rough boundary is undefined, its partial dimensions can be associated %-- instead of directions -- 
more directly to the maximal and cross-boundary Laypunov exponents, both of which can be defined without a reference to directions, in an operational manner. {\bfb The latter is important insomuch that we can check if the condition $\lambda_x<\lambda_U$ for roughness is satisfied.} The cross-boundary LE $\lambda_x$, {\bfac on the one hand}, can be defined by the divergence of ensembles initialised {\bfac according to the natural measure of the nonattracting set,} on the two sides very near the boundary, say, in terms of the ensemble means\footnote{It can be defined also in terms of, say, the Wasserstein distance (see~\cite{Robin2017393} for an application), but computationally this is much less efficient or robust.}. The MLE $\lambda_U$, {\bfac on the other hand}, can also be measured by the divergence of trajectory-pairs (as done in~\cite{0951-7715-30-7-R32}). The difference between these definitions of $\lambda_x$ and %the MLE 
$\lambda_{U}$ is the opposite order of A) {\bfac measuring} the difference and B) {\bfac taking the} ensemble averages wrt. the natural measure supported by the nonatracting set. 

%Since in the climate model studied in~\cite{0951-7715-30-7-R32} the climatic and atmospheric weather time scales are vastly separated, due to eq. (\ref{eq:uncert}) the co-dimension of the boundary is $D-D_{b,0}\approx0$, and so it was not clear whether the measured $1-D_1^{(c)}\approx0$ was $\alpha=D-D_{b,0}\approx0$ or rather  $1-D_0^{(x)}=0$. We demonstrated here that the nontrivial $1-D_1^{(c)}=\alpha$ can indeed be measured by the said technique. 

\begin{figure*} [t!]
    \begin{center}
            \scalebox{0.5}{\includegraphics{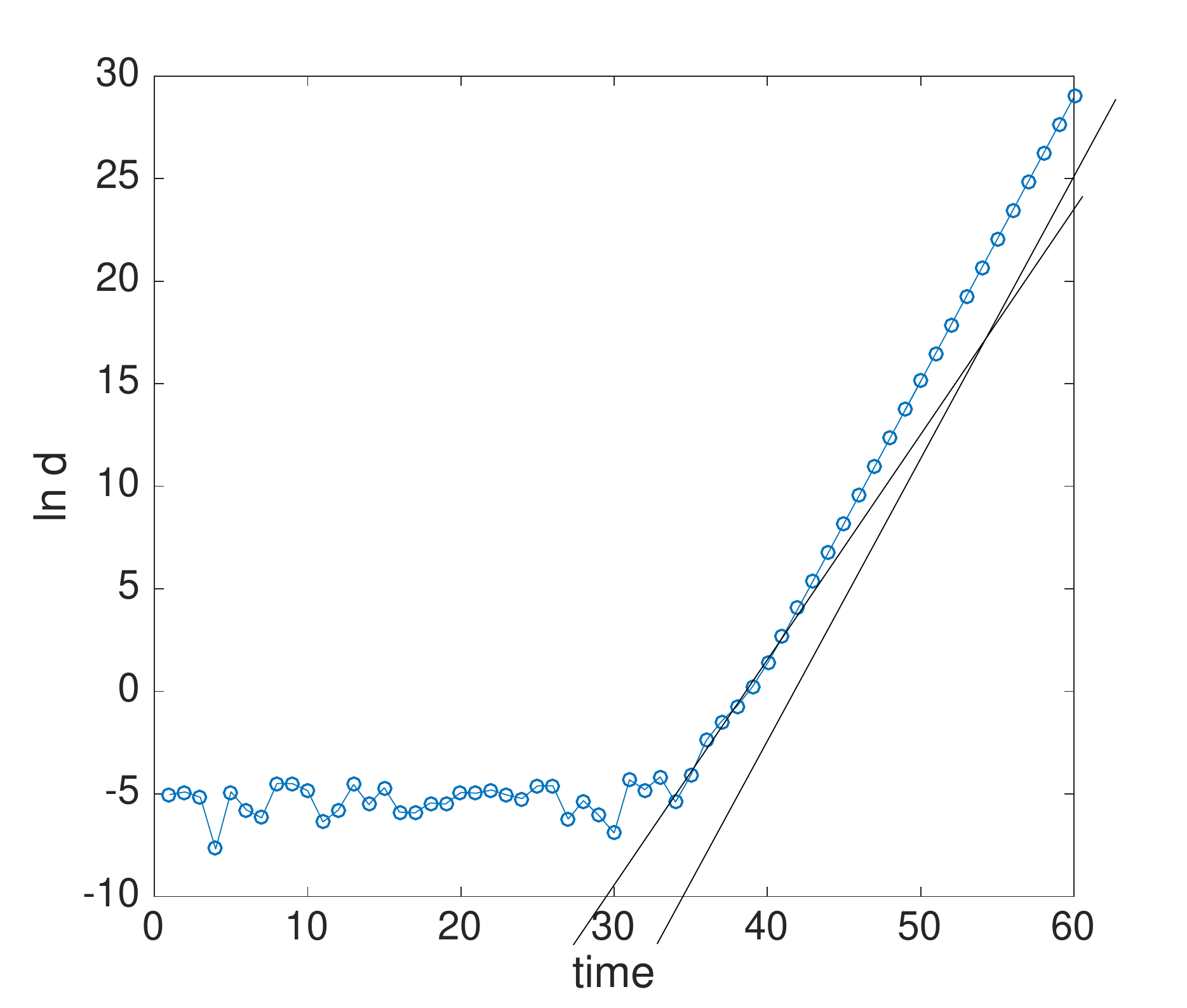}} 
        \caption{\label{fig:lambda_x} 
        Distance of ensemble means as time progresses, where the two ensembles comprise trajectories with opposite outcomes, as shown in Fig. \ref{fig:sprinkler}. Straight black lines of slopes $\ln 4\approx 1.39$ and $\lambda_x=1.18$ aid the interpretation of the diagram.}
    \end{center}
\end{figure*}

{\bfb We are not aware of a method by which the formal results of $D_1^{(x)}=1$ and $D_1^{(y)}<1$ can be verified. By traversing the basin boundary with a line and evaluating the uncertainty exponent of the resulting intersection set, owing to eq. (\ref{eq:dim_intersect}), it is not necessarily the partial dimension $D_1^{(x)}$ that we can measure, but the always nontrivial co-dimension $D-D_{b,1}$.}
Therefore, from a practical point of view, as far as the predictability of the second kind alone is concerned, it makes no difference if the basin boundary is a filamentary or a rough fractal. Conversely, we cannot identify the type of fractal by measuring fractal dimensions. We conclude that it can only be identified by measuring both the maximal Lyapunov exponent, as done in~\cite{0951-7715-30-7-R32}, and the cross-boundary Lyapunov exponent, upon which the theoretical condition (Sec. \ref{sec:condition}) can be checked. %The cross-boundary LE can be approximated simply by measuring the exponential separation (in phase space) of ensemble means where the two ensembles consist of trajectories with opposite outcomes. % 11.01.2020. it's just above now
We have done this for our prototype model (\ref{eq:mixed_2})-(\ref{eq:mixed_1}) featuring a new type of rough boundary, using the ensemble shown in Fig. \ref{fig:sprinkler}, and show the results in Fig. \ref{fig:lambda_x}. Given that the ensembles are of finite size, initially no exponential separation can be seen. The separation rate yielded by large values of $\ln d$, that is, when the ensembles departed substantially from the nonattracting set, is $\ln a$, and it clearly does not pertain to the nonattracting set for which eq. (\ref{eq:mixed_LEs}) predicts the correct value to be $\lambda_x\approx1.18$. We drew a line of that slope into the diagram and find that the initial separation rate (at small values of $\ln d$) is clearly consistent with it. Therefore, the technique proposed for identifying the type of fractal seems to be viable. While a visual identification is possible only in 2D or perhaps 3D systems, the computation of the two LEs should not be hindered by large dimensionality.

We also emphasize that our prediction of the dimension (\ref{eq:part_dim_y}) assumes small perturbations; but the condition for roughness has no such assumption. {\bfb Otherwise, our new eq. (\ref{eq:uncert}), the generalization of the Kantz-Grassberger relation (KG), indicates that the ratio of $\lambda_x$ and $\lambda_U$ not only determines the type of the fractal, but has a bearing also on its dimension. These LEs associated respectively with weather and climatic process in our model in~\cite{0951-7715-30-7-R32} resulted in an almost completely space filling basin boundary, i.e., a complete unpredictability of the outcome on fine scales.}

Because of the inapplicability of KG, {\bfb point (ii) of Sec.~\ref{sec:Intro}} is not universally valid. Indeed the local and global instabilities can coincide {\bfb [compare eqs. (\ref{eq:uncert}) and (\ref{eq:dim_pred_reg})]}, i.e., the fractality due to roughness does not necessarily weaken global instability (in proportion with the co-dimension) like in the case of filamentary fractals. However, we note that when e.g. the roughening happens through perturbation, there could be a weakening of global instability according to the effect first reported in~\cite{Franaszek:1991} and also examined e.g. in~\cite{AE:2010}, whereby the maximal effect occurs for some nontrivial perturbation strength $\epsilon_X$.
%The systems in which such a boundary occurs, we view such that two subsystems are coupled mutually, where one subsystem is responsible for the bistability, such that without coupling it is readily bistable, and the other susbystem -- without coupling -- performs some `persistent' dynamics possessing an attractor. Directions in the full phase space that are given by the variables of the respective subsystems we will refer to as `unstable direction' and `persystent direction'. Lyapunov exponents and partial dimensions can then be associated with these directions. Despite that (ii) applies in these systems, it turns out, from the present contribution, that the escape takes place not in the unstable but the persistent direction. Therefore, it will be necessary to {\em distinguish between an unstable direction and a `direction of escape'}. In the case of filamentary (continuous) fractal boundary the direction of escape is the unstable (persistent) direction. 
In contrast, (i.a) remains valid when the boundary is readily rough (when the $X$ subsystem itself maybe a coupled/perturbed system). (i.b), on the other hand, is not necessarily valid, as roughening the boundary by perturbation might change only the fractal dimension of the boundary but not the global instability, as already said. When the global instability is altered by perturbation according to~\cite{Franaszek:1991} and~\cite{AE:2010}, however, it should contribute to `further' change in the dimension similarly to filamentary fractals~\cite{PhysRevE.87.042902}. The latter, of course, falls under the validity of (i.a), when the parameter in question can be taken to be the perturbation strength, $p=\epsilon_X$.

The decoupled ocean dynamics in~\cite{0951-7715-30-7-R32} was given by a diffusive energy balance model, the Ghil-Sellers model~\cite{Ghil:1976}, studied also in~\cite{Bodai2015}, whose solution is not chaotic, and so the fractality of the basin boundary in the coupled model came only from roughness. In a state-of-the-art Earth system model, or even in an intermediate complexity model with a dynamical ocean like the Planet Simulator atmospheric model coupled to the Goldstein ocean-sea ice model~\cite{gmd-9-3347-2016}, we expect a mixed filamentary-rough fractal basin boundary to be present. It is also clear that due to the vast atmospheric vs climatic time scale separation the basin boundary is practically space filling. However, the answers to two related questions of geophysical relevance are not immediately clear:

\begin{itemize}
 \item With a realistic Earth-like setup, at what length scale (in phase space) can we resolve the space-filling roughness? % contribution $\hat{D}_1$ to the fractal dimension would the roughness have?
 \item Is the atmospheric perturbation strong enough to significantly alter the global instability as in~\cite{Franaszek:1991,AE:2010,PhysRevE.87.042902}?
\end{itemize}

% 14.01.2020.
{\bfc In a different approach, atmospheric perturbations of the slow climatic model components can be viewed as noise perturbations, possibly inducing transitions or exits from one persistent state to another. Key questions concern the most likely path of transition as well as the expected residence time in the persistent state~\cite{PhysRevLett.122.158701,LB:2020}. These properties are governed by a potential-like quantity~\cite{LT:2011}. An efficient algorithm is proposed in Ref.~\cite{Bodai:2020} to estimate the potential barrier height from controlled exit time data.}

\section*{Acknowledgments}

This work was supported financially from the EU Horizon 2020 grant for project CRESCENDO (under grant no. 641816). VL thanks Bruno Eckhardt, Celso Grebogi, Arkady Pikovsky and James Yorke for useful discussions. {\bfac The authors are grateful for Tam\'as T\'el for his feedback on an earlier version of the manuscript and for discussions during ``The Mathematics of Climate and the Environment'' programme of the Institut Henri Poincar\'e.%, where this manuscript was finalised.
}

% \appendix
% 
% \section*{Appendix: Prediction skill with a discrete precursor}\label{sec:binary_precursor}
% 
% We might not have an appendix

%\bibliographystyle{Copernicus}
%\bibliographystyle{plain}
\bibliographystyle{unsrt}

\bibliography{bb_global_instab_08}

\end{document}